\RequirePackage{fix-cm}
\documentclass[sn-mathphys]{sn-jnl}

\usepackage{lmodern}
\usepackage{amsmath}
\usepackage{caption}
\usepackage{esvect}
\usepackage{subcaption}
\usepackage{units}
\usepackage{xcolor}
\usepackage{makecell}
\setcellgapes{1pt}
\usepackage{algorithm}
\usepackage{algpseudocode}
\usepackage[capitalise,noabbrev]{cleveref}
\usepackage{array}   

\newcolumntype{C}{>{\centering\arraybackslash$}p{1.1207cm}<{$}}
\algnewcommand{\LineComment}[1]{\State \(\triangleright\) #1}

\jyear{2024}

\title[Resolving social dilemmas with minimal reward transfer]{Resolving social dilemmas with minimal reward transfer}

\author*[1]{\fnm{Richard} \sur{Willis} }
\email{richard.willis@kcl.ac.uk}

\author*[1]{\fnm{Yali} \sur{Du}}
\email{yali.du@kcl.ac.uk}

\author[1,2]{\fnm{Joel Z.} \sur{Leibo}}
\email{jzl@deepmind.com}

\author[3]{\fnm{Michael} \sur{Luck}}
\email{michael.luck@sussex.ac.uk}

\affil*[1]{
\orgname{King's College London}, \orgaddress{
\city{London},
\country{United Kingdom}}}

\affil[2]{\orgname{Google DeepMind}, \orgaddress{\city{London}, \country{United Kingdom}}}

\affil[3]{\orgname{University of Sussex}, \orgaddress{\city{Sussex}, \country{United Kingdom}}}

\begin{document}

\abstract{
Social dilemmas present a significant challenge in multi-agent cooperation because individuals are incentivised to behave in ways that undermine socially optimal outcomes. Consequently, self-interested agents often avoid collective behaviour.
In response, we formalise social dilemmas and introduce a novel metric, the \textit{general self-interest level}, to quantify the disparity between individual and group rationality in such scenarios.
This metric represents the maximum proportion of their individual rewards that agents can retain while ensuring that a social welfare optimum becomes a dominant strategy.
Our approach diverges from traditional concepts of altruism, instead focusing on strategic reward redistribution.
By transferring rewards among agents in a manner that aligns individual and group incentives, rational agents will maximise collective welfare while pursuing their own interests.
We provide an algorithm to compute efficient transfer structures for an arbitrary number of agents, and introduce novel multi-player social dilemma games to illustrate the effectiveness of our method.
This work provides both a descriptive tool for analysing social dilemmas and a prescriptive solution for resolving them via efficient reward transfer contracts.
Applications include mechanism design, where we can assess the impact on collaborative behaviour of modifications to models of environments.
}


\maketitle

\section*{Acknowledgement}
This work was supported by UK Research and Innovation [grant number EP/S023356/1], in the UKRI Centre for Doctoral Training in Safe and Trusted Artificial Intelligence (\url{www.safeandtrustedai.org}) and a BT/EPSRC funded iCASE Studentship [grant number EP/T517380/1].

\section{Introduction}
\label{sec:Introduction}
Social dilemmas are situations in which individuals face the choice between acting selfishly for personal gain or acting in a prosocial manner to benefit the collective, which yields greater overall benefits.
The problem with social dilemmas is that it does not pay to be nice to others; individuals are incentivised to behave in ways that undermine socially optimal outcomes.
Consequently, self-interested agents avoid collective action, resulting in suboptimal outcomes for everyone.
Although each agent prefers mutual collective behaviour over mutual selfish behaviour, they are individually powerless to bring this about. 
This tension between collective and individual rationality is characterised by agents who engage in selfish behaviour outperforming those engaged in collective behaviour within a group, while prosocial groups outperform selfish groups.

For example, suppose a group of villagers has unrestricted access to a forest, and each villager can profit from logging trees.
The number of trees that can be harvested at a sustainable level is limited, however, and the forest will degrade if excessive logging takes place.
The collective benefits most when the forest is managed sustainably, maximising the number of trees that can be logged over time, but the villagers individually profit from each additional tree they harvest.
This particular social dilemma represents a common pool resource, and is known as a \emph{tragedy of the commons}.

Social dilemmas are significant because they occur in many diverse situations, and there are still instances of failures to enact collective action in the world today.
Real-world social dilemmas include the routine administration of antibiotics to farm animals, which benefits the farmer but increases antibiotic resistance, negatively impacting society.
Similarly, in file sharing networks, where users can either act for personal gain by downloading files or help the network by uploading files, there exist so called \emph{free-riders} who do not contribute.
Consequently, methods to resolve such dilemmas, achieved by causing all agents to prefer to take collective action, are important.

While humans have developed biases, norms, institutions and other mechanisms that help them to cooperate in such situations~\cite{ostrom92__covenants_with_and_without_a_sword,dietz03__the_struggle_to_govern_the_commons}, it is unclear whether artificial agents will possess these same behaviours.
As the prevalence of artificial agents continues to rise, finding solutions applicable to them becomes increasingly important.
The algorithms used by artificial agents typically make decisions based on logic and mathematics, so we believe there is a need for solutions that apply to instrumentally rational agents.

Many solutions to social dilemmas have been developed~\cite{apt14__selfishness_level_of_strategic_games,deng18__disarmament_games_with_resources,nowak06__five_rules_for_the_evolution_of_cooperation,hughes18__inequity_aversion_improves_cooperation_in_intertemporal_social_dilemmas,hughes20__learning_to_resolve_alliance_dilemmas_in_manyplayer_zero-sum_games}, but they typically have specific requirements to be applicable, such as relying on agents to have prosocial preferences, which limits their applicability.
We argue that a solution should require only a single agreement between participants and be scalable to an arbitrary number of agents.
It should also rely solely on the incentives present in the scenario and not require agents to have private knowledge of, or previous experience with others.
To our knowledge, only a potential application of trading of reward shares~\cite{schmid23__learning_to_participate_through_trading_of_reward_shares} could meet these criteria.
Rather than buying a stake in the future rewards of another agent, which requires a valuation of the fair value of such a deal, our method involves agents voluntarily committing to transfer proportions of their future rewards to one another in a manner that benefits all participants.

In this paper, we represent social dilemmas abstractly as games, which are analysed using game theory, and we propose a solution that resolves these dilemmas while addressing the issues highlighted above.
The key principle underpinning our approach is that if we sufficiently align individual and group incentives, rational agents maximising their own rewards will also maximise group rewards.
To align individual incentives, we introduce the concept of agents committing to share proportions of their future game rewards with others, a technique that we label \emph{reward transfer}.
By engaging in reward transfer, an agent provides incentives for the recipients to help it prosper, which, paradoxically, can lead to a net profit for the transferring agent if it causes a beneficial behavioural change in the recipients.

We investigate two specific instances of reward transfer.
First, we consider when all agents exchange the same proportion of their reward between one another.
We show that a sufficiently large amount of reward transfer from all agents is guaranteed to resolve a social dilemma, and we determine the minimum proportion that must be exchanged to do so.
Second, we investigate the general case, where each player may transfer any amount of their reward to any other player, and introduce a method to find the minimum transfer arrangement that will resolve a social dilemma.
By determining the maximum amount of their own reward that agents may keep for themselves while resolving a social dilemma, we quantify, for a given game, the discrepancy between what is rational for the individual compared to the collective.
If, in order for prosocial actions to be rational, individuals can only retain a small amount of self-interest, then cooperation is challenging in this game.
Conversely, if agents can keep a large proportion of their own rewards, needing only to transfer small proportions to others, then the game is conducive to collective action.
This provides our work with a descriptive element.
By determining the minimal amount of reward that must be transferred to others, agents can use our results as an efficient, prescriptive solution to social dilemmas by entering the proposed transfers contract, which makes cooperation rational for all.

Our analytical findings reveal that superficially similar multi-player social dilemmas can exhibit dramatically different cooperative incentives.
We show that the manner in which the agents influence one another affects how the difficulty of incentivising cooperation varies with the number of participants.
Furthermore, the effectiveness of resolving these dilemmas depends on the permitted structure of reward transfers.
In some cases, when an appropriate transfer structure is in place, the difficulty of resolving social dilemmas does not necessarily increase as the number of participants rises.
However, in other scenarios, achieving cooperation among many participants may require them to share almost all of their rewards.

Concretely, we make the following contributions: we propose a novel formalisation of social dilemmas and what it means to resolve them (\cref{sec:social_dilemma}); we introduce the \emph{symmetrical self-interest level}, a metric quantifying the minimum proportion of future rewards that must be exchanged by all agents to resolve a social dilemma (\cref{sec:exchange}); and we introduce the \emph{general self-interest level}, a second, more general metric representing the maximum self-interest agents can retain while using reward transfers to resolve a dilemma (\cref{sec:general_transfer}).
Additionally, we present an algorithm to find an optimal reward transfer structure among agents and analyse its computational complexity (\cref{sec:method_algorithm}); we construct novel multi-player social dilemma games using graphs (\cref{sec:game_setup}); and we provide parameterised results for the games to illustrate how the challenge of incentivising cooperation depends on the shape of the graph and number of players (\cref{sec:results}).

The paper is structured as follows.
In \cref{sec:related_work} we review existing work on game metrics and solutions to facilitating collective action in social dilemmas.
Our main contribution is developed in \cref{sec:method}, and we provide results in \cref{sec:experiments}.
Finally, we conclude with a discussion in \cref{sec:conclusion}.

\section{Related work}
\label{sec:related_work}
We begin by comparing the descriptive element of our work to other game theoretical metrics that quantify properties of a game in \cref{sec:metrics}.
The remainder of the discussion explores mechanisms to achieve collective action in social dilemmas and positions our work in relation to them.
We start with norms and strategic play (\cref{sec:norms,sec:strategy}), which are behavioural solutions that discourage antisocial behaviour, often by punishing offenders.
We then deal with approaches which lead learning algorithms to cooperate in social dilemmas via the use of intrinsic rewards in \cref{sec:shaping}.
Next, we consider how games can be extended with notions of commitments in \cref{sec:contracts}.
\cref{sec:rw_reward_transfer} discusses extending games with reward transfer, the mechanism we use in our work.
The final \cref{sec:rw_commitment_mechanisms} provides an overview of commitment implementations, discussing how agents enter binding agreements in practice.

\subsection{Metrics}
\label{sec:metrics}
Metrics are used to quantify and describe features of a game.
Certain metrics specify how difficult it is to achieve a particular outcome.
For example, \emph{K-implementation}~\cite{deng16__complexity_and_algorithms_of_kimplementation} determines the cost for a third party who provides additional incentives in order to make a particular outcome dominant.
The \emph{selfishness level} of a game measures the willingness of the players to cooperate~\cite{apt14__selfishness_level_of_strategic_games}; it is the smallest fraction of the social welfare of each outcome that needs to be offered to each player so that a social optimum is realised in a pure Nash equilibrium.
Our work is closely related, but our solution concept is more general, as we can, in effect, offer different amounts to different players within the same game outcome, potentially finding more efficient solutions.
The selfishness level uses a model of altruism based on~\cite{elias10__sociallyaware_network_design_games} and finds equivalent results for the models of altruism proposed in~\cite{chen08__altruism_selfishness_and_spite_in_traffic_routing,chen11__the_robust_price_of_anarchy_of_altruistic_games,caragiannis10__the_impact_of_altruism_on_the_efficiency_of_atomic_congestion_games}.
Though we do not base our solution concept on altruism, our simpler metric, the symmetrical self-interest level, has a similar form to that of Chen and Kempe~\cite{chen08__altruism_selfishness_and_spite_in_traffic_routing}.

Other metrics quantify the price a group of agents pays when they seek a particular solution concept instead of a social optima.
The Price of Anarchy~\cite{koutsoupias09__worstcase_equilibria}, Price of Stability~\cite{anshelevich04__the_price_of_stability_for_network_design_with_fair_cost_allocation} and Price of Pareto Optimality~\cite{elkind20__price_of_pareto_optimality_in_hedonic_games} measure the ratio between the best Nash equilibrium, the worst Nash equilibrium, or the worst Pareto optimal outcome (respectively) and a social welfare optimum.
Although these metrics assess the quality of certain game outcomes, they say little about how such outcomes are achieved.
In our work, we assess the difficulty of maximising social welfare, rather than quantifying the loss if one is not achieved.

\subsection{Norms}
\label{sec:norms}
Societies often have enforced collective expectations of behaviour.
For example, most communities expect their members to dispose of refuse appropriately, and anyone littering will likely attract social sanctions, such as being called out.
These expected behaviours are called \emph{norms}.
Successful norms are self-sustaining because it is beneficial to follow them, or because the agents internalise the behaviours.
A norm can be supported by \emph{punishment}, whereby any agent who violates the norm is penalised.
Norms can be viewed as a soft constraint: an agent is not forced to undertake the behaviour prescribed by a norm, but they usually will because it is typically in their best interests to do so.

Axelrod~\cite{axelrod86__an_evolutionary_approach_to_norms}, and Mahmoud et al.~\cite{mahmoud10__an_analysis_of_norm_emergence_in_axelrods_model} investigate when a norm can be evolutionarily stable.
Montes and Sierra~\cite{montes21__valueguided_synthesis_of_parametric_normative_systems,sierra21__value_alignment} propose a method to determine the most effective norms to promote a given set of values, while Han~\cite{han22__institutional_incentives_for_the_evolution_of_committed_cooperation} explores how institutional incentives can most effectively promote cooperation when commitments are costly.
Researchers experimentally found norms that lead to the reciprocity of cooperation~\cite{ohtsuki06__the_leading_eight} and investigated the societal effects when agents adhere to guilt and fairness norms~\cite{pereira17__social_manifestation_of_guilt_leads_to_stable_cooperation_in_multiagent_systems,lorini18__the_longterm_benefits_of_following_fairness_norms_under_dynamics_of_learning_and_evolution}.
Limitations of norms include the fact that they are not always beneficial to society, such as norms around gender roles, and they can be difficult to introduce and establish.
We are interested in solutions to social dilemmas that are not dependent on shared behavioural expectations, so that they can be immediately utilised by novel groups of agents.

\subsection{Strategic play}
\label{sec:strategy}
Repeated games and games involving a temporal component can permit cooperative strategies if they are enforceable via punishment~\cite{jacq20__foolproof_cooperative_learning}.
In the famous example, \emph{Tit-For-Tat}~\cite{axelrod80__effective_choice_in_the_prisoners_dilemma} plays the action its partner chose in the previous round, thus reciprocating cooperation while minimising any exploitation from a selfish partner.
In a two-player repeated social dilemma game, a strategy can be robust to exploitation attempts if it only recalls the result of the last round~\cite{press12__iterated_prisoners_dilemma_contains_strategies_that_dominate_any_evolutionary_opponent}.
In these circumstances, a strategy can stabilise cooperation with a partner by ensuring that their opponent does best under mutual cooperation, and that it is costly for their opponent to outperform them~\cite{hilbe13__evolution_of_extortion_in_iterated_prisoners_dilemma_games,stewart13__from_extortion_to_generosity_evolution_in_the_iterated_prisoners_dilemma}.
Unfortunately, it can be difficult to detect when an opponent is shirking cooperation, and players have an incentive to do so if they can avoid punishment.
Our approach uses incentives instead of punishments to support cooperation.

The success of a strategy depends on the other strategies it encounters.
It can be useful to consider how the composition of strategies in a population will change over time, using techniques such as evolutionary game theory or replicator dynamics~\cite{hofbauer03__evolutionary_game_dynamics,nowak04__evolutionary_dynamics_of_biological_games}.
With these techniques, it is possible to investigate conditions that can lead to the emergence of cooperation in a population~\cite{nowak06__five_rules_for_the_evolution_of_cooperation}.
Unfortunately, the existence of punishment actions does not guarantee that it will be used in support of cooperation, as selfish agents may punish cooperators to prevent them from succeeding~\cite{rand11__the_evolution_of_antisocial_punishment_in_optional_public_goods_games}.

Due to backwards induction, cooperation can often only be a Nash equilibrium in social dilemmas if games are infinitely repeated, and if they are, then according to the \emph{folk theorem}~\cite{leyton-brown08__essentials_of_game_theory}, there become infinitely many Nash equilibria, which can lead to players disagreeing over their preferred equilibria.
Consequently, strategic play can add cooperative equilibria because it increases the range of strategies available to the players, but it does not remove equilibria that lead to poor outcomes.
Our solution is effective in zero-shot single-round games and avoids the need for punishment.

\subsection{Reward shaping}
\label{sec:shaping}
A rational, self-interested agent only cares about the rewards it receives, but agent designers can encourage certain behaviours by including auxiliary intrinsic rewards in addition to the game reward.
In reinforcement learning literature, this is known as \emph{reward shaping}.

When an intrinsic reward is proportional to the mean collective reward~\cite{peysakhovich17__prosocial_learning_agents_solve_generalized_stag_hunts_better_than_selfish_ones}, it is regarded as a \emph{prosocial reward}, and has been shown to help reinforcement learning agents converge to a better equilibrium in a social dilemma called Stag Hunt (\cref{table:sd_c}).
McKee et al.~\cite{mckee20__social_diversity_and_social_preferences_in_mixedmotive_reinforcement_learning} generate different degrees of prosocial rewards in mixed-motive games; they train populations with homogeneous and heterogeneous social preferences and find some evidence that heterogeneous populations achieve more equal payoffs.

According to Haeri~\cite{haeri21__rewardsharing_relational_networks_in_multi-agent_reinforcement_learning_as_a_framework_for_emergent_behavior}, an agent can be considered to have a relationship with another specific agent if it has a personal reward term that depends on the reward of the other agent.
In the general case, a weighted graph can be used to specify the relationships between all agents; 
Haeri explores how different relationship networks lead to differences in performance.
Rather than specifying intrinsic rewards by hand, they can be determined via an optimisation process, such as using a genetic algorithm to optimise a reward shaping neural network~\cite{wang19__evolving_intrinsic_motivations_for_altruistic_behavior}.

Alternatively, we can give agents a negative intrinsic reward when they achieve lower than average rewards, which creates an incentive for the group to prefer egalitarian rewards~\cite{hughes18__inequity_aversion_improves_cooperation_in_intertemporal_social_dilemmas}.
This can be an effective approach in social dilemmas when agents have access to a punishment mechanism: should an agent benefit from selfish behaviour, the other agents will receive lower than average rewards, and their intrinsic reward motivates them to punish the selfish agent.
In this way, punishment helps to prevent agents from learning selfish behaviour.

While reward shaping has proven effective at promoting collective behaviour in social dilemmas, it often does not explain where the additional rewards come from, relying on an innate disposition of the agents themselves.
By adjusting behaviour to include factors beyond the game reward, intrinsic motivations may not be rational to adopt, making them unsuitable for a group of self-interested, independent agents.
Our approach uses rational, voluntary transfers of the extrinsic game reward to promote prosocial behaviour.

\subsection{Commitments and contracts}
\label{sec:contracts}
In social dilemmas, the key question is how to ensure mutual cooperation.
One option is to use \emph{binding agreements}, or contracts, between players, which are known as \emph{cooperative games} in game theory literature.
Han~\cite{han17__evolution_of_commitment_and_level_of_participation_in_public_goods_games} and Ogbo et al.~\cite{ogbo22__evolution_of_coordination_in_pairwise_and_multiplayer_interactions_via_prior_commitments} show that participation commitments facilitate better outcomes in public goods dilemmas by compensating those who make costly prosocial choices.
Hughes et al.~\cite{hughes20__learning_to_resolve_alliance_dilemmas_in_manyplayer_zero-sum_games} demonstrate how extending a stochastic game to include a protocol for agreeing joint action contracts improves outcomes.
Our approach also uses contracts as a mechanism for resolving social dilemmas, but our agreements involve only transfers of rewards, rather than mandating actions or behaviours, which can be more challenging to specify or enforce.

If contracts can include a \emph{side payment}, then an agent can be paid to take an action that benefits another.
Building on this, Christoffersten et al.~\cite{christoffersen22__get_it_in_writing} formalise contracts for stochastic games to include payments.
While it isn't clear how much should be transferred, a fixed point, called the Coco value~\cite{sodomka13__cocoq}, has been proposed around which players of a two-player game could negotiate the fair value of a joint action.
Although joint action contracts provide assurance of behaviour, they can be burdensome to specify:
Not only must an agent be able to specify the contracts it would enter for all possible joint actions in all possible states, but if it is costly to write a contract, agents will incur a cost each time step they enter one.

Instead of a joint commitment to play certain actions, Deng and Conitzer~\cite{deng17__disarmament_games,deng18__disarmament_games_with_resources} propose that a player could alternatively commit unilaterally to avoid playing certain (possibly mixed) strategies, a method called \emph{disarmament}.
By removing all non-prosocial actions, or at least reducing the probability that an agent may play them, better social outcomes can be achieved.
Although the authors introduce a negotiation protocol that leads to improvements in the expected welfare of the participants, disarmament cannot resolve all social dilemmas.

\subsection{Reward transfer}
\label{sec:rw_reward_transfer}
Reward transfer is a technique whereby agents send and receive rewards from one another in a zero-sum manner.
An agent may choose to transfer rewards to another to influence the recipient's behaviour in a way that benefits the transferor.
In economics, the task of designing an employment contract to optimally motivate an employee is known as the \emph{principal-agent problem}~\cite{lambert86__executive_effort_and_selection_of_risky_projects,demski87__delegated_expertise,malcomson09__principal_and_expert_agent}.
More recently, Markov games have been extended to include reward transfers between agents.
Explicitly adding an action that allows a player to gift rewards to their peers has been explored in a tragedy of the commons scenario~\cite{lupu20__gifting_in_multiagent_reinforcement_learning} and in coordination games~\cite{wang21__emergent_prosociality_in_multiagent_games_through_gifting}.
Yang et al.~\cite{yang20__learning_to_incentivize_other_learning_agents} optimise the transferred reward to shape the behaviour of learning algorithms, thereby improving the overall reward to the gifting agent.
Similar to the limitations on contracts (\cref{sec:contracts}), if rewards are transferred at the granularity of individual actions, then a strategy must include the transfers it would make for all joint actions for all states, which is burdensome.
Our solution is simpler in that this decision is made only once, before the game is played.

An alternative to specifying payments for actions is to take a stake in the future outcomes of an agent.
Baker et al.~\cite{baker20__emergent_reciprocity_and_team_formation_from_randomized_uncertain_social_preferences} use reward transfer to represent varying inter-agent social preferences and investigate how it impacts teamwork.
Yi et al.~\cite{yi22__learning_to_share_in_multiagent_reinforcement_learning} enable reinforcement learning agents to dynamically exchange rewards with those closest to them, which helps to achieve collective behaviour.
Gemp et al.~\cite{gemp22__d3c} develop an algorithm for learning algorithms in stochastic games that reduces the Price of Anarchy through reward transfers (here called \emph{loss-sharing}).
During training, the players learn a loss-sharing matrix that increases the social welfare of the worst-case local Nash equilibrium solution.
A marketplace that enables the trading of reward shares has been suggested by Schmid et al.~\cite{schmid23__learning_to_participate_through_trading_of_reward_shares}. 
Their approach is most similar to ours, but we offer a simpler method involving players choosing to donate their rewards rather than acquiring a right to the rewards of other players.
By determining the limiting amount of shared interest required, our work incorporates a descriptive element, and, should reward transfer be costly, minimises such cost by identifying the smallest amount required to be transferred.

\subsection{Commitment mechanisms}
\label{sec:rw_commitment_mechanisms}
Our work proposes an optimal structure of reward transfers that agents should commit to.
The implementation of this commitment is left open, but we provide some examples below.
In certain situations, actors can be trusted to comply with their promises because societal pressures provide strong incentives for them to maintain a trustworthy reputation or follow conventions, as discussed in \cref{sec:norms}.
However, this is not always the case, so concrete methods to ensure compliance are necessary.
There are a range of existing methods for enforcing joint commitments, and which is most suitable depends on the particular scenario in question.

Legal contracts are appropriate for governing real-world interactions between humans or corporations.
These use third-party enforcement mechanisms to ensure compliance.
Should a party renege on their contractual obligation, their counterpart can seek recourse through legal proceedings.
Smart contracts~\cite{taherdoost23__smart_contracts_in_blockchain_technology}, widely used in blockchain, are composed of computer code and execute when predetermined conditions are met.
They are most suitable for supply chain or financial services, frequently being used within cryptocurrencies to ensure terms without the use of a third-party.
As AI progresses, it may enable new ways to make joint commitments~\cite{conitzer23__foundations_of_cooperative_ai}.
One proposal is that artificial agents submit computer programs to interact on their behalf, a paradigm known as \emph{program games}~\cite{tennenholtz04__program_equilibrium,oesterheld19__robust_program_equilibrium}.
These programs, or delegates, can inspect or simulate~\cite{kovarik23__game_theory_with_simulation_of_other_players} the other programs in advance to verify that their co-players will perform the agreed behaviours.

\section{Methodology}
\label{sec:method}
We review normal-form games and formalise the notion of a social dilemma in \cref{sec:nfg,sec:social_dilemma}.
In \cref{sec:reward_transfer}, we introduce our reward transfer concept and present two metrics which describe properties of a social dilemma game in \cref{sec:exchange,sec:general_transfer}.
\cref{sec:method_algorithm} proposed an algorithm to compute the metric introduced in the previous section, the general self-interest level.
We conclude with a discussion in \cref{sec:method_discussion}.

\subsection{Preliminaries}
\label{sec:nfg}
The normal-form game, also known as the strategic or matrix form, is a representation of every player’s utility in each state of the world, for the special case where states of the world depend only on the players’ combined actions.
A (finite, \emph{n}-person) normal-form game is a tuple $(N, A, \vv{R})$, where:
\begin{itemize}
    \item $N$ is a finite set of $n$ players, indexed by $i$.
    \item $A = A_1 \times ... \times A_n$, where $A_i$ is a finite set of actions available to player $i$. Each tuple $\vv{a} = (a_1 , ..., a_n) \in A$ is called an action profile.
    \item $\vv{R} = (R_1 , ... , R_n)$ where $R_i : A \rightarrow \mathbb{R}$ is a real-valued reward (or payoff) function for player $i$.
\end{itemize}

Normal-form games are naturally represented by an \emph{n}-dimensional matrix.
An action is a player's best response to the fixed action profile of their opponent(s) if it maximises their reward.
If an action is the best response to all possible opponent action profiles, then we say that it is \emph{dominant}.
In what follows, for convenience, we write $\vv{a_{-i}}$ to represent the tuple of actions of all players other than player $i$.
An action profile $\vv{a}$ is a \emph{Nash equilibrium} if, for all agents $i$, $a_i$ is a best response to $\vv{a_{-i}}$.
An action profile is \emph{dominant} if, for all agents $i$, $a_i$ is the best response to all possible opponent action tuples.

We introduce the notion of collective good, represented by a \emph{social welfare metric}, $SW : \mathbb{R}^n \rightarrow \mathbb{R}$, indicating the overall benefit to a group.
An action profile is a \emph{social welfare optimum} if it maximises the social welfare metric.
In this paper, we use the utilitarian metric, $U$, as our social welfare metric.
The utilitarian metric measures the unweighted sum of rewards obtained by all players, as follows.

\begin{equation}
\label{eqn:utilitarian}
U(\vv{r}) = \sum_{i=1}^n r_i
\end{equation}

Where $\vv{r} = (r_1, ..., r_n)$, is a tuple of individual rewards.

\subsection{Social dilemmas}
\label{sec:social_dilemma}
We begin by formalising what it means for a game to be a social dilemma in order to demonstrate how our proposed solution directly addresses their core challenges.
We describe a new formalism for what constitutes a social dilemma and what it means to resolve one.
At the end of this section, we mention alternative existing formalisations from literature, and in \cref{sec:method_discussion} we discuss the advantages of our characterisation.

A social dilemma is a game in which all players can choose to take prosocial actions (to cooperate) at a potential cost to themselves.
While the benefit to the group is higher when players cooperate, the benefit to an individual player may be greater when they act selfishly (to defect).
Therefore, players face a choice between acting for the benefit of the group or in their own interests.

A social dilemma can therefore be defined as a situation in which, for all agents: (i) the social welfare of the group is strictly greater when an agent chooses to cooperate than when it chooses to defect, regardless of the actions of the other agents; and (ii) each agent does better individually when it defects.
We also include the condition (iii) that all agents prefer the action profile of mutual cooperation over mutual defection.
This reflects the fact that the social welfare increase from mutual cooperation is distributed widely, rather than being captured by only a subset of agents.

Consider a general-sum normal-form game where each agent faces a choice to either cooperate, $C$, or defect, $D$.
We define $^\frown$ as a coupling operation that inserts $a_i$ into $\vv{a_{-i}}$ such that \(\vv{a} = a_i^\frown \vv{a_{-i}}\).
For a given social welfare metric, $SW(\vv{r})$, we define a strict social dilemma as follows.
For any action profile $\vv{a} \in A$:

\begin{align*}
\text{(i)} \quad & \forall i \quad SW(R(C^\frown \vv{a_{-i}})) > SW(R(D^\frown \vv{a_{-i}}))\\
\text{(ii)} \quad & \forall i \quad R_i(D^\frown \vv{a_{-i}}) > R_i(C^\frown \vv{a_{-i}})\\
\text{(iii)} \quad & \forall i \quad R_i((C,C,...,C)) > R_i((D,D,...,D))
\end{align*}

A \emph{strict} social dilemma is characterised by a unique, pure Nash equilibrium in which all players defect, and this equilibrium is both a social welfare minimum and Pareto inferior, meaning that there exists at least one outcome where all players are better off.
This follows from the fact that defection is a dominant strategy for every player, the social welfare strictly decreases with each defection, and everyone prefers the mutual cooperation action profile.

A \emph{partial} social dilemma occurs when each agent may benefit from defecting, depending on the choices of their opponent(s).
In this respect, there exists at least one action profile $\vv{a_{-i}}$ that would lead an agent to prefer defection.
Thus, while there is always a personal cost to cooperation in a strict social dilemma, this is not always the case in a partial social dilemma.
For a partial social dilemma, the second inequality above can therefore be softened as follows. 
\begin{equation}
\nonumber
\text{(ii)} \quad \forall i\ \; \exists \vv{a_{-i}} : R_i(D^\frown \vv{a_{-i}}) > R_i(C^\frown \vv{a_{-i}})
\end{equation}

\cref{table:sd} shows three matrix games that are commonly considered social dilemmas in the literature~\cite{macy02__learning_dynamics_in_social_dilemmas,leibo17__multiagent_reinforcement_learning_in_sequential_social_dilemmas}.
Under our definition, Prisoner's Dilemma (\cref{table:sd_a}) is an example of a strict social dilemma, while Chicken (\cref{table:sd_b}) and Stag Hunt (\cref{table:sd_c}) are examples of partial social dilemmas.

\begin{table}[t]
  \centering
 \begin{subtable}{0.3\linewidth}
    \centering
    \begin{tabular}{c|cc}
      & $C$ & $D$\\
      \hline
      $C$ & $3,3$ & $0,4$\\
      $D$ & $4,0$ & $1,1$\\
    \end{tabular}
    \subcaption{\label{table:sd_a}Prisoner's Dilemma}
  \end{subtable}
  \begin{subtable}{0.3\linewidth}
    \centering
    \begin{tabular}{c|cc}
      & $C$ & $D$\\
      \hline
      $C$ & $3,3$ & $1,4$\\
      $D$ & $4,1$ & $0,0$\\
    \end{tabular}
    \subcaption{\label{table:sd_b}Chicken}
  \end{subtable}
  \begin{subtable}{0.3\linewidth}
    \centering
    \begin{tabular}{c|cc}
      & $C$ & $D$\\
      \hline
      $C$ & $4,4$ & $0,3$\\
      $D$ & $3,0$ & $1,1$\\
    \end{tabular}
    \subcaption{\label{table:sd_c}Stag Hunt}
  \end{subtable}
  \caption{\label{table:sd}Normal-form social dilemmas}
\end{table}

We say that a social dilemma is \emph{resolved} if an action profile that maximises social welfare is dominant.
Rational players will always take a dominant action, so this solution concept guarantees that a social welfare optimum will occur.

A normal-form social dilemma remains a social dilemma under scalar translation and positive scalar multiplication of the rewards.
Different social welfare metrics admit different games.
For example, John Rawls proposed the Maximin principle \cite{rawls71__a_theory_of_justice}: that we should maximise the welfare of the worst-off members of society.
A metric representing this principle would take the minimum reward of all players.
Under this metric, Chicken is still a partial social dilemma, but Prisoner's Dilemma and Stag Hunt are not, because our definition excludes games where cooperation may decrease social welfare.

While others have previously defined social dilemmas, we believe our formalisation, which focuses on costly prosocial choices, better represents the underlying dynamics of these dilemmas.
Macy and Flache~\cite{macy02__learning_dynamics_in_social_dilemmas} consider them to be normal-form games with payoffs satisfying certain conditions, but some of these conditions are justified only if players hold particular beliefs.
In turn, Hughes et al.~\cite{hughes18__inequity_aversion_improves_cooperation_in_intertemporal_social_dilemmas} draw on the work of Schelling~\cite{schelling73__hockey_helmets_concealed_weapons_and_daylight_saving} and generalise the Macy and Flache definition to apply to \emph{n}-player Markov games.
However, both Hughes et al.\cite{hughes18__inequity_aversion_improves_cooperation_in_intertemporal_social_dilemmas} and Schelling~\cite{schelling73__hockey_helmets_concealed_weapons_and_daylight_saving} include within their definition games with social optima that occur when a number of players defect.
Similarly, previous works have used alternative notions of resolving a social dilemma, such as the selfishness level of a game~\cite{apt14__selfishness_level_of_strategic_games} which defines it as when a social optimum is realised in a pure Nash equilibrium.
We return to these alternative definitions of a social dilemma and what it means to resolve them in \cref{sec:method_discussion}, after we have introduced our mechanism.

\subsection{Reward transfer}
\label{sec:reward_transfer}
As discussed above, the primary difficulty of social dilemmas is that prosocial actions are personally costly.
Agents need sufficient motivation to care about others for collective action to become more attractive than selfish behaviour.
To address this, an agent must provide some form of incentive for others to care about its well-being, which it can achieve by sharing its rewards.

Consider the villagers in our tragedy of the commons example in \cref{sec:Introduction}.
Suppose the villagers enter into a contract to donate a proportion of the wood they harvest to be shared equally among the other villagers.
Now, each villager has a reduced incentive to harvest more trees than is sustainable, for two reasons.
First, they only get to keep a fraction of their own proceeds, as they must share their profits with others.
Second, if they refrain from behaviour that degrades the forest, there will be more trees harvested in the future, and the individual in question will be entitled to a proportion of the increased yields that the other villagers obtain.
If this effective tax level is correctly set, then the villagers will refrain from unsustainable logging.

Thus, in this section, we introduce a mechanism by which agents can commit to transferring proportions of their future rewards to one another.
This mechanism requires the existence of \emph{transferable utility}, a common currency equally valued by all players, and the ability for the agents to enter into \emph{binding agreements} (see \cref{sec:rw_commitment_mechanisms} for possible implementations).
We consider two cases: first, where a group of agents enters a contract to exchange equal proportions of their future rewards, and second, the general case, where each agent is free to specify different proportions of their future rewards to commit to transfer to particular agents.
In what follows, our examples involve games with non-negative rewards, but negative game rewards, or losses, are transferred in the same way.

\subsection{Reward exchange}
\label{sec:exchange}
Formally, when all players enter into a contract to retain a proportion $s \in [0,1]$ of their future rewards, and to divide the remainder equally amongst the other agents, we call this concept \emph{reward exchange}.
In this way, each agent receives a proportion $\frac{1-s}{n-1}$ of the future rewards of all other agents.
The \emph{post-transfer reward} to agent $i$, $r_i'$, is the retained part of their individual game reward, $r_i$, plus any reward received from others, and is given by:

\begin{eqnarray}
\label{eqn:reward_exchange}
r_i'(\vv{r}, s) & = & s r_i + \frac{1-s}{n-1}\sum_{j \neq i} r_j
\end{eqnarray}

Consider again Prisoner's Dilemma (\cref{table:sd_a}), in which the players agree to retain $s = \frac{3}{4}$ of their payoffs, as shown in \cref{table:re_a}.
Here, both players are ambivalent between cooperating or defecting.
This occurs because each player stands to gain only $s$ as much from a possible defect action, and cooperating increases the reward of their opponent, of which they are entitled to a proportion $(1-s)$.
If the players retain an amount $s < \frac{3}{4}$, cooperation is dominant for both, and the dilemma is resolved.
However, for any value of $s > \frac{3}{4}$, the players retain too much self-interest and stand to gain from defection, so the dilemma remains unresolved.
In Chicken and Stag Hunt, cooperation is weakly dominant for both players when $s = \frac{2}{3}$ (\cref{table:re_b,table:re_c}).

We call the limiting value of $s$ that resolves the dilemma the \emph{symmetrical self-interest level} of a game, denoted \(s^*\), because each player retains the same self-interest in their own reward and exchanges an equal amount with all other players.
In \cref{sec:general_transfer} we will generalise the transfers by relaxing the symmetrical nature.

\begin{table}[t]
  \centering
  \begin{subtable}{0.4\linewidth}
    \centering
    \begin{tabular}{c|cc}
      & $C$ & $D$\\
      \hline
      $C$ & $3,3$ & $1,3$\\
      $D$ & $3,1$ & $1,1$\\
    \end{tabular}
    \caption{Prisoner's Dilemma, $s^*=\frac{3}{4}$}
    \label{table:re_a}
  \end{subtable}
  \begin{subtable}{0.28\linewidth}
    \centering
    \begin{tabular}{c|cc}
      & $C$ & $D$\\
      \hline
      $C$ & $3,3$ & $2,3$\\
      $D$ & $3,2$ & $0,0$\\
    \end{tabular}
    \caption{Chicken, $s^*=\frac{2}{3}$}
    \label{table:re_b}
  \end{subtable}
  \begin{subtable}{0.3\linewidth}
    \centering
    \begin{tabular}{c|cc}
      & $C$ & $D$\\
      \hline
      $C$ & $4,4$ & $1,2$\\
      $D$ & $2,1$ & $1,1$\\
    \end{tabular}
    \caption{Stag Hunt, $s^*=\frac{2}{3}$}
    \label{table:re_c}
  \end{subtable}
  \caption{\label{table:re}Symmetrical self-interest level of the social dilemmas}
\end{table}

Formally, the symmetrical self-interest level of a game is the maximum proportion of their own reward $s$ that players can retain while resolving the social dilemma.
Writing the reward function returning the post-transfer reward to player $i$ as $R'_i(\vv{r}, s)$, we have:

\begin{equation}
\label{eqn:symmetrical_self_interest_level}
s^* = \max \{s \mid \forall i \; R'_i(C^\frown \vv{a_{-i}}, s) > R'_i(D^\frown \vv{a_{-i}}, s)\}
\end{equation}

The symmetrical self-interest level of a game is invariant to scalar translation and positive scalar multiplication of the game rewards, and exists for all social dilemmas under the utilitarian metric (Equation~\eqref{eqn:utilitarian}).
This follows from the fact that when every player retains $s = \frac{1}{n}$ of their rewards, Equation~\eqref{eqn:reward_exchange} reduces to:

\begin{equation}
\label{eqn:max_exchange_minimum}
r_i'(\vv{r},\frac{1}{n}) = \frac{1}{n} \sum_{j=1}^n{r_j} = \frac{1}{n} U(\vv{r})
\end{equation}

Consequently, each player seeks to maximise the utilitarian metric, and since cooperation increases social welfare, there is a unique social welfare maximum of mutual cooperation.
The symmetrical self-interest level, therefore, has a lower bound, and $s^* \in [\frac{1}{n},1]$.
We therefore expect that, all else being equal, increasing the number of players in a game will decrease its symmetrical self-interest level.

The symmetrical self-interest level of a game can be viewed as a measure of the players' willingness to cooperate; it quantifies the disparity between individual and group incentives.
A high symmetrical self-interest level indicates that players need to care only a little about each other to achieve a socially optimal outcome, whereas a low symmetrical self-interest level suggests that players have strong incentives to shirk prosocial behaviour.
Importantly, if we can determine the symmetrical self-interest level, we may also be able to find a way to resolve social dilemmas through agents exchanging a proportion $\frac{1-s}{n-1}$ of their future rewards with one another.

We note that the symmetrical self-interest level of finitely iterated normal-form games is the same as for a single-round game.
This is due to backward induction, whereby a player realises that in the final round they will play as they would in a single-round game.
Consequently, in the penultimate round, because the strategies of the players will not impact their play in the final round, there is no reason to deviate from the best single round strategy here either.
This logical reasoning continues until the current round.
In order to ensure that mutual cooperation is dominant, therefore, sufficient reward must be exchanged such that players prefer to cooperate in the current round.

Reward exchange is similar to the concept of \emph{prosocial} rewards from Peysakhovich and Lerer~\cite{peysakhovich17__prosocial_learning_agents_solve_generalized_stag_hunts_better_than_selfish_ones}, a form of reward shaping (\cref{sec:shaping}) in which an agent modifies only its own reward by including an additional intrinsic reward term. 
However, our notion of reward transfer involves only a redistribution of the \emph{extrinsic} game rewards and allows an agent to impact the rewards of other agents (in addition to its own reward).
Reward exchange is also similar to the model of altruism proposed by Chen and Kempe~\cite{chen08__altruism_selfishness_and_spite_in_traffic_routing}, which can be thought of as agents transferring a proportion of their rewards into a common fund, which is then distributed equally amongst all agents.
In their mechanism, therefore, an agent will receive back a proportion $\frac{1}{n}$ of the rewards it transferred to the fund.
The difference between our approaches will be relevant in the next section.

\subsection{Generalised reward transfer}
\label{sec:general_transfer}
So far, we have assumed that each player divides their relinquished reward equally amongst their co-players.
This, in effect, measures how much of the combined group reward the players need in order to act for the collective.
A benefit of this approach is its simplicity, requiring only a single parameter.
However, it may be possible to be more efficient and find ways to reduce the amount of future rewards that must be transferred to others by exploiting the game structure.
For example, in a situation that is spatially located, agents may only be impacted by their local neighbourhood, needing only to share rewards with those closest to them to incentivise cooperation.
In this case, transferring rewards to agents that are further away may be unnecessary.

We now examine the general case of reward transfer and define the \emph{reward transfer matrix}, $\mathbf{T}$, which specifies the proportion of their game reward that the row player transfers to the column player.
The tuple of post-transfer rewards, $\vv{r'}$, is given by the game rewards, $\vv{r}$, multiplied by the reward transfer matrix, $\vv{r'} = \mathbf{T} \vv{r}$.

\begin{equation}
\label{eqn:rtm}
  \mathbf{T} =
  \begin{vmatrix}
    t_{1,1} & t_{1,2} & \cdots & t_{1,n} \\
    t_{2,1} & t_{2,2} & \cdots & t_{2,n} \\
    \vdots & \vdots & \ddots & \vdots \\
    t_{n,1} & t_{n,2} & \cdots & t_{n,n}
  \end{vmatrix}
\end{equation}

Where $\forall i \; \forall j \; t_{ij} \in [0,1]$, meaning that a player can transfer at most all their reward to another player and cannot transfer a negative proportion.
Furthermore, we require that the rows sum one, $\forall i \; \sum_j{t_{ij}} = 1$, ensuring that players account for all of their reward, thereby conserving the total game reward.
We do not, therefore, permit the players to engage in \emph{utility burning}~\cite{moulin76__cooperation_in_mixed_equilibrium}.
The columns need not sum to one, however, so any reward-conserving distribution is permissible.
Reward exchange is therefore a specific case of generalised reward transfer using a symmetrical reward transfer matrix of the following form.

\begin{equation}
\nonumber
  \mathbf{T} =
  \begin{vmatrix}
    s & \frac{1-s}{n-1} & \cdots & \frac{1-s}{n-1} \\
    \frac{1-s}{n-1} & s & \cdots & \frac{1-s}{n-1} \\
    \vdots & \vdots & \ddots & \vdots \\
    \frac{1-s}{n-1} & \frac{1-s}{n-1} & \cdots & s
  \end{vmatrix}
\end{equation}

Exactly as with reward exchange, we are interested in the maximum proportion of their own rewards that each player may retain while resolving a social dilemma.
We refer to the diagonal values, $t_{ii}$, of the reward transfer matrix as the self-interest of the players.
Out of all possible reward transfer matrices that make cooperation dominant for all players, we find the transfer matrix with the largest minimum of diagonal elements.
This is the matrix with the greatest amount of self-interest that the agent(s) with the least self-interest retain, and we call the value of its minimum diagonal element the \emph{general self-interest level} of the game, denoted by $g^*$.
Formally, writing $diag(\mathbf{T})$ as a function returning the tuple of diagonal values of $\mathbf{T}$, and the reward function returning the post-transfer reward to player $i$ as $R'_i(\vv{r}, \mathbf{T})$, we have.

\begin{equation}
\label{eqn:general_self_interest_level}
g^* = \max \{\min(diag(\mathbf{T})) \mid \forall i \; R'_i(C^\frown \vv{a_{-i}}, \mathbf{T}) > R'_i(D^\frown \vv{a_{-i}}, \mathbf{T})\}
\end{equation}

We refer to a reward transfer matrix that achieves the general self-interest level as a \emph{minimal reward transfer matrix}, and we denote it as $\mathbf{T}^*$.
As demonstrated in Equation \eqref{eqn:max_exchange_minimum}, the symmetrical self-interest level is guaranteed to exist for a social dilemma under the utilitarian metric.
The general self-interest level is therefore also guaranteed to exist, because we can always use the reward transfer matrix equivalent to the symmetrical self-interest level.
However, we may be able to find a greater value for the general self-interest level.
The general self-interest level of a game is likewise invariant under scalar translation and positive scalar multiplication of the game rewards.
We now illustrate this metric with two possible three-player variants of Prisoner's Dilemma.

Previously, we examined a two-player Prisoner's Dilemma (\cref{table:sd_a}) and found the symmetrical self-interest level to be $s^* = \frac{3}{4}$.
The payoffs for this game can be constructed using the following rules: a player gains a reward of 1 when they defect and gives a reward of 3 to their opponent when they cooperate.
Using similar rules, we create two three-player variants.
In both of the variants, a player who defects gains a reward of 1, as before.
In the first version (\cref{table:s_pd}), cooperation now divides the reward of 3 equally between both co-players, giving each $\frac{3}{2}$; we call this variant Symmetrical-3PD.
In the second version (\cref{table:c_pd}), cooperate gives a reward of 3 to the player with index \(i + 1 \mod n\), which we call Cyclical-3PD.
In this latter case, cooperation only increases the reward of \emph{one} other agent.

\begin{table}[ht]
  \centering
  \caption{\label{table:s_pd}Symmetrical-3PD}
  \begin{subtable}{0.49\linewidth}
    \centering
    \makegapedcells
    \begin{tabular}{c|cc}
      & $C$ & $D$\\
      \hline
      $C$ & $3,3,3$ & $\frac{3}{2},4,\frac{3}{2}$\\
      $D$ & $4,\frac{3}{2},\frac{3}{2}$ & $\frac{5}{2},\frac{5}{2},0$\\
    \end{tabular}
    \subcaption{Player 3 cooperates}
  \end{subtable}
  \begin{subtable}{0.49\linewidth}
    \centering
    \makegapedcells
    \begin{tabular}{c|cc}
      & $C$ & $D$\\
      \hline
      $C$ & $\frac{3}{2},\frac{3}{2},4$ & $0,\frac{5}{2},\frac{5}{2}$\\
      $D$ & $\frac{5}{2},0,\frac{5}{2}$ & $1,1,1$\\
    \end{tabular}
    \subcaption{Player 3 defects}
  \end{subtable}
\end{table}

\begin{table}[ht]
  \centering
  \caption{\label{table:c_pd}Cyclical-3PD}
  \begin{subtable}{0.49\linewidth}
    \centering
    \begin{tabular}{c|cc}
      & $C$ & $D$\\
      \hline
      $C$ & $3,3,3$ & $3,4,0$\\
      $D$ & $4,0,3$ & $4,1,0$\\
    \end{tabular}
    \subcaption{Player 3 cooperates}
  \end{subtable}
  \begin{subtable}{0.49\linewidth}
    \centering
    \begin{tabular}{c|cc}
      & $C$ & $D$\\
      \hline
      $C$ & $0,3,4$ & $0,4,1$\\
      $D$ & $1,0,4$ & $1,1,1$\\
    \end{tabular}
    \subcaption{Player 3 defects}
  \end{subtable}
\end{table}

While the symmetrical self-interest level for both these games is $s^* = \frac{3}{5}$, the general self-interest level of Symmetrical-3PD is $g^* = \frac{3}{5}$ compared to $g^* = \frac{3}{4}$ for Cyclical-3PD.
The minimal reward transfer matrices are, respectively:

\begin{equation}
\nonumber
\mathbf{T}^* = 
\begin{vmatrix}
\nicefrac{3}{5} & \nicefrac{1}{5} & \nicefrac{1}{5}\\
\nicefrac{1}{5} & \nicefrac{3}{5} & \nicefrac{1}{5}\\
\nicefrac{1}{5} & \nicefrac{1}{5} & \nicefrac{3}{5}
\end{vmatrix}
\text{ and }
\mathbf{T}^* = 
\begin{vmatrix}
\nicefrac{3}{4} & 0 & \nicefrac{1}{4}\\
\nicefrac{1}{4} & \nicefrac{3}{4} & 0\\
0 & \nicefrac{1}{4} & \nicefrac{3}{4}
\end{vmatrix}
\end{equation}

In Cyclical-3PD, the minimal reward transfer matrix permits the players to retain a larger proportion of their own rewards compared to the symmetrical self-interest level.
This is because the rewards for each player depend only on their own action and the action of one other player.
Consequently, each player only needs to offer a proportion of their rewards to the player who impacts their game reward.
The situation is different for Symmetrical-3PD, where each player must incentivise both co-players to cooperate, resulting in a greater general self-interest level.
Note that for this game, the minimal reward transfer matrix is not unique, and we are free to choose the off-diagonal elements to be any value within bounds as long as the columns and rows all sum to one.
Post-reward transfers, each player gains $\nicefrac{3}{5}$ when they defect.
They need to benefit at least as much when they cooperate, which means receiving at least $\nicefrac{2}{5}$ of the rewards of their co-players.
This only happens if all columns sum to one.
We generalise these games and find their symmetrical and general self-interest levels as a function of the number of players in \cref{sec:experiments}.

\subsection{Algorithm}
\label{sec:method_algorithm}
Given a normal-form social dilemma (\cref{sec:social_dilemma}), we can use a linear program to find a reward transfer matrix that retains the greatest proportion of self-interest for the players while resolving the dilemma.
Our objective is to maximise the minimum of the diagonals in the reward transfer matrix, $\mathbf{T}$.
We introduce an auxiliary variable, $z$, to convert the non-linear objective function (due to the minimum) into a linear objective function.
Writing the $i$-th column in $\mathbf{T}$ as $\mathbf{T}_{\cdot i}$, our linear program is as follows.

\begin{align*}
  \max_{z, \mathbf{T}} \; \quad   & z \\
  \text{subject to: }  \quad  & t_{ii} \geq z,   \quad && \text{(i)} \\
  &  0 \leq t_{ij} \leq 1, \quad &&  \text{(ii)} \\
  & \sum_{k=1}^n t_{ik} = 1, \quad &&  \text{(iii)} \\
  &  \mathbf{T}_{\cdot i} \cdot \mathbf{r}(D^\frown \mathbf{a}_{-i}) \leq \mathbf{T}_{\cdot i} \cdot \mathbf{r}(C^\frown \mathbf{a}_{-i}), \quad &&  \text{(iv)} \\[1.5ex]
  & \forall i,j \in\{1, \ldots, n\}, \forall \mathbf{a}_{-i} \in \{C,D\}^{n-1}.
\end{align*}

Constraints (i) are the $n$ auxiliary constraints.
In addition, we have (ii) bounds on the $n^2$ elements of $\mathbf{T}$  and (iii) $n$ equality constraints that each row of $\mathbf{T}$ sums to one.
We resolve the dilemma by ensuring that there are no action profiles where a player prefers to defect after rewards are redistributed.
Essentially, wherever a player has an incentive to defect, their co-players compensate them sufficiently for cooperating, thereby making mutual cooperation dominant.
We achieve this by formulating (iv) inequality constraints that govern the parameters of $\mathbf{T}$.
For each of the $n$ players, we have one inequality for each of the $2^{n-1}$ possible co-player action profiles, giving us $2^{n-1} n$ constraints.

\cref{alg:mgsd} describes a program to find the minimal reward transfer matrix for a social dilemma using a linear program, such as the interior-point method~\cite{potra00__interiorpoint_methods}.
Depending on the choice of linear program algorithm, the solver could have a worst-case performance that grows either parametrically or exponentially with the number of parameters or constraints~\cite{brand20__a_deterministic_linear_program_solver_in_current_matrix_multiplication_time}.
In practice, many of the game reward constraints may be redundant, and linear programs typically achieve far better performance than their worst-case guarantees~\cite{megiddo87__on_the_complexity_of_linear_programming}.
However, because it takes exponential time to traverse the $2^n$ action profiles and generate $2^{n-1}n$ inequality constraints, the memory and computational complexity of \cref{alg:mgsd} is exponential.
The memory requirements of representing a game with $2^n$ action profiles is already exponential, however, so this result is not surprising.

We demonstrate and verify our algorithm on two games and provide an indicative time to solve for a range of values of $n$ in \cref{sec:algorithm_results}.
Note that for games that are symmetrical under permutation, such as Cyclical-3PD, each row in the reward transfer matrix is a permutation of the first row.
In this simpler case, we can avoid the use of an auxiliary variable and only need to find the $n$ parameters of a row.

\begin{algorithm}[tb]
\caption{An algorithm to compute the minimal reward transfer matrix}
\label{alg:mgsd}
\begin{algorithmic}
\Require linear program, $LP$
\Require normal-form social dilemma, $N$, with $n$ players
\Require set of all possible action profiles, $A$
\State $auxillary$ $variable$, $z$
\State $reward$ $transfer$ $matrix$, $T = matrix(n$ by $n)$
\State $bounds = list(t_{ij} \in [0,1]$ for $i$ in $1...n$ for $j$ in $1...n)$
\State $constraints = list()$

\For{$i$ in $1...n$}
    \State $constraints \gets t_{ii} > z$
    \Comment{auxiliary constraints}
    \State $constraints \gets \sum_j t_{ij} = 1$
    \Comment{rows sum to one}
\EndFor

\For{each $action \text{ } profile, \vv{a} \in A$}
  \For{$i$ in $1...n$}
    \If{$\vv{a}[i] = C$}
      \LineComment{compute the change in reward for all players if player $i$ plays D}
      \LineComment{player $i$ must not benefit from defection after reward transfers}
      \State $constraints \gets T[:,i] \cdot (N[D ^\frown \vv{a_{-i}}] - N[C ^\frown \vv{a_{-i}}]) \leq 0$
    \EndIf
  \EndFor
\EndFor

\State $LP($objective: maximise $z$, parameters: $(z, T)$, $bounds$, $constraints)$
\end{algorithmic}
\end{algorithm}

\subsection{Discussion}
\label{sec:method_discussion}
We now return to our discussion from \cref{sec:social_dilemma} regarding the formalisation of a social dilemma, and the definition of what it means to resolve one.
The formalisation proposed by Hughes et al.~\cite{hughes18__inequity_aversion_improves_cooperation_in_intertemporal_social_dilemmas} includes games where cooperation is not guaranteed to increase social welfare.
We argue that while such games can manifest elements of a social dilemma, these complex games are better viewed as a combination of a social dilemma and another game in which players benefit from coordinating their actions, such as Volunteer's Dilemma (\cref{table:volunteer}) or Pure Coordination (\cref{table:coordination}).
Here, the payoff table shows the reward to the row player given the joint action of all players.

\begin{table}[ht]
  \centering
  \caption{Coordination games}
  \begin{subtable}{0.49\linewidth}
    \centering
    \begin{tabular}{c|cc}
      & \makecell{At least 2 players\\cooperate} & Otherwise\\
      \hline
      $C$ & $0$ & $-1$\\
      $D$ & $2$ & $-1$\\
    \end{tabular}
    \subcaption{Volunteer's Dilemma\label{table:volunteer}}
  \end{subtable}
  \begin{subtable}{0.49\linewidth}
    \centering
    \begin{tabular}{c|cc}
      & \makecell{All players take\\the same action} & Otherwise\\
      \hline
      $C$ & $1$ & $0$\\
      $D$ & $1$ & $0$\\
    \end{tabular}
    \subcaption{Pure Coordination\label{table:coordination}}
  \end{subtable}
\end{table}

To illustrate the differences, we introduce a game called \emph{Too Many Cooks in Prison}, displayed in \cref{table:cooks}.
This game features the payoffs of Symmetrical-3PD (\cref{table:s_pd}) minus those of the Pure Coordination (\cref{table:coordination}).
One interpretation of this game could be individuals in society deciding whether to comply with tax laws (cooperate) or practice tax evasion (defect).
The government can fund public goods, which are more efficient at delivering utility to society than individual spending.
However, while there are certain high-value projects that the government will preferentially deliver, there are decreasing returns to these projects, so if the government receives too much tax revenue, it will waste the money on white elephant projects.

\begin{table}[ht]
  \centering
  \caption{\label{table:cooks}Too Many Cooks in Prison}
  \begin{subtable}{0.49\linewidth}
    \centering
    \makegapedcells
    \begin{tabular}{c|cc}
      & $C$ & $D$\\
      \hline
      $C$ & $2,2,2$ & $\frac{3}{2},4,\frac{3}{2}$\\
      $D$ & $4,\frac{3}{2},\frac{3}{2}$ & $\frac{5}{2},\frac{5}{2},0$\\
    \end{tabular}
    \subcaption{Player 3 cooperates}
  \end{subtable}
  \begin{subtable}{0.49\linewidth}
    \centering
    \makegapedcells
    \begin{tabular}{c|cc}
      & $C$ & $D$\\
      \hline
      $C$ & $\frac{3}{2},\frac{3}{2},4$ & $0,\frac{5}{2},\frac{5}{2}$\\
      $D$ & $\frac{5}{2},0,\frac{5}{2}$ & $0,0,0$\\
    \end{tabular}
    \subcaption{Player 3 defects}
  \end{subtable}
\end{table}

For this game, social welfare is maximised when only two out of three players cooperate.
Hence, this is not a social dilemma under our definition, while it is under that of Hughes et al.~\cite{hughes18__inequity_aversion_improves_cooperation_in_intertemporal_social_dilemmas}.
Nevertheless, there are strong individual incentives to shirk prosocial actions, and defect is a weakly dominant strategy in this game, so the players need incentives to cooperate.
Though providing incentives to players to desire a social welfare optimum is necessary, it is not sufficient to ensure such an optimum is realised, because they also need to coordinate on a particular action profile.
Too Many Cooks in Prison is a symmetrical game, so it is not possible to use reward exchange (\cref{sec:exchange}) to make a social welfare optimum a dominant action profile, because reward exchange does not alter this symmetry.
This is also the case for the models of altruism in  literature~\cite{elias10__sociallyaware_network_design_games,chen08__altruism_selfishness_and_spite_in_traffic_routing,chen11__the_robust_price_of_anarchy_of_altruistic_games,caragiannis10__the_impact_of_altruism_on_the_efficiency_of_atomic_congestion_games}.
\cref{table:cooks_exchange} shows the game using reward exchange (Equation~\eqref{eqn:reward_exchange}) $s=\frac{3}{5}$, which is the limiting value that makes the social optima of $(D,C,C)$, $(C,D,C)$ and $(C,C,D)$ all Nash equilibria.

\begin{table}[ht]
  \centering
  \caption{\label{table:cooks_exchange}Too Many Cooks in Prison with reward exchange}
  \begin{subtable}{0.49\linewidth}
    \centering
    \makegapedcells
    \begin{tabular}{c|cc}
      & $C$ & $D$\\
      \hline
      $C$ & $2,2,2$ & $2,3,2$\\
      $D$ & $3,2,2$ & $2,2,1$\\
    \end{tabular}
    \subcaption{Player 3 cooperates}
  \end{subtable}
  \begin{subtable}{0.49\linewidth}
    \centering
    \makegapedcells
    \begin{tabular}{c|cc}
      & $C$ & $D$\\
      \hline
      $C$ & $2,2,3$ & $1,2,2$\\
      $D$ & $2,1,2$ & $0,0,0$\\
    \end{tabular}
    \subcaption{Player 3 defects}
  \end{subtable}
\end{table}

Under the notion of resolving a social dilemma provided by Apt and Schafer~\cite{apt14__selfishness_level_of_strategic_games}, this would qualify as there is a social welfare-maximising pure Nash equilibrium.
However, this is insufficient to guarantee that the played action profile will maximise social welfare for two reasons.
First, if there are multiple pure social welfare-maximising Nash equilibria, as is the case here, the players face an equilibrium selection problem.
If they are not able to coordinate, they do not know which of the social optima they should aim for.
Second, even if there were only one pure Nash equilibrium, if it is not dominant, the players may still fail to take their appropriate action if they have reason to believe that other player(s) may deviate from their action.

Generalised reward transfer (\cref{sec:general_transfer}) can be used to make a social welfare optimum dominant if the players choose which action profile to coordinate on in advance.
\cref{table:cooks_resolved} shows Too Many Cooks in Prison using the following minimal reward transfer matrix, which makes the action profile $\vv{a} = (D,C,C)$ dominant.

\begin{equation}
\nonumber
  \mathbf{T}^* =
  \begin{vmatrix}
    \frac{3}{11} & \frac{4}{11} & \frac{4}{11} \\
    0 & \frac{3}{11} & \frac{8}{11} \\
    0 & \frac{8}{11} & \frac{3}{11}
  \end{vmatrix}
\end{equation}

\begin{table}[ht]
  \centering
  \caption{\label{table:cooks_resolved}Too Many Cooks in Prison resolved using reward transfer}
  \begin{subtable}{0.49\linewidth}
    \centering
    \makegapedcells
    \begin{tabular}{c|cc}
      & $C$ & $D$\\
      \hline
      $C$ & $\frac{6}{11},\frac{30}{11},\frac{30}{11}$ & $\frac{9}{22},\frac{30}{11},\frac{85}{22}$\\
      $D$ & $\frac{12}{11},\frac{65}{22},\frac{65}{22}$ & $\frac{15}{22},\frac{35}{22},\frac{30}{11}$\\
    \end{tabular}
    \subcaption{Player 3 cooperates}
  \end{subtable}
  \begin{subtable}{0.49\linewidth}
    \centering
    \makegapedcells
    \begin{tabular}{c|cc}
      & $C$ & $D$\\
      \hline
      $C$ & $\frac{9}{22},\frac{85}{22},\frac{30}{11}$ & $0,\frac{5}{2},\frac{5}{2}$\\
      $D$ & $\frac{15}{22},\frac{30}{11},\frac{35}{22}$ & $0,0,0$\\
    \end{tabular}
    \subcaption{Player 3 defects}
  \end{subtable}
\end{table}

Users should therefore specify which solution concept is appropriate for resolving a social dilemma.
For example, if the players have access to a coordinating device, and can play a correlated equilibrium~\cite{leyton-brown08__essentials_of_game_theory}, the symmetrical self-interest level of the game would be $s=\frac{3}{5}$, and reward exchange at such a level would be an appropriate solution.
In the general case, without access to a correlating signal, only general reward transfer is appropriate, and the minimal reward transfer matrix gives $g^* = \frac{3}{11}$.

An alternative definition of social dilemmas is provided by Kollock~\cite{kollock98__social_dilemmas}, who defines social dilemmas as games with Pareto inferior Nash equilibria, and calls them \emph{severe} if they have a Pareto inferior dominant action profile.
While we acknowledge that Pareto inferior Nash equilibria are challenging and symptomatic of social dilemmas, we contend that this definition is overly broad.
As an illustrative example, Pure Coordination (\cref{table:coordination}) with two players has a weak Nash equilibrium where both players play C or D with equal probability, yielding an expected payoff of one half.
While this action profile leaves both players worse off than they could otherwise be, it stems from an inability to coordinate, rather than from conflicting incentives: in fact, the players value the action profiles identically.

Social dilemmas are intrinsically connected to notions of fairness.
The general self-interest level is entirely focused on guaranteeing social welfare optima, regardless of the fairness of the reward distribution in such an action profile.
For illustration, consider the game of Prisoner's Dilemma where the rewards to player 1 in all action profiles have been multiplied by a factor of 3, and an infinitesimal amount $\epsilon$ has been subtracted from their unilateral defection action profile, so that cooperation increases social welfare and the game conforms to our definition of a social dilemma.
We call this game Scaled-PD, and it is displayed in \cref{table:scaled_pd_a}.

\begin{table}[ht]
  \centering
  \caption{\label{table:scaled_pd}Scaled-PD}
  \begin{subtable}[t]{0.32\linewidth}
    \centering
    \makegapedcells
    \flushleft
    \begin{tabular}[t]{c|cc}
      & $C$ & $D$\\
      \hline
      $C$ & $9,3$ & $0,4$\\
      $D$ & $12-\epsilon,0$ & $3,1$\\
    \end{tabular}
    \subcaption{No transfers\label{table:scaled_pd_a}}
  \end{subtable}
  \begin{subtable}[t]{0.32\linewidth}
    \centering
    \makegapedcells
    \begin{tabular}[t]{c|cc}
      & $C$ & $D$\\
      \hline
      $C$ & $6,6$ & $2,2$\\
      $D$ & $\frac{6-\epsilon}{2},\frac{6-\epsilon}{2}$ & $2,2$\\
    \end{tabular}
    \subcaption{\label{table:scaled_pd_b}With $\mathbf{T}^* = \begin{vmatrix}\nicefrac{1}{2} & \nicefrac{1}{2}\\\nicefrac{1}{2} & \nicefrac{1}{2}
\end{vmatrix}$}
  \end{subtable}
  \begin{subtable}[t]{0.32\linewidth}
    \centering
    \makegapedcells
    \flushright
    \begin{tabular}[t]{c|cc}
      & $C$ & $D$\\
      \hline
      $C$ & $6,6$ & $0,4$\\
      $D$ & $8-\frac{2\epsilon}{3},4-\frac{\epsilon}{3}$ & $2,2$\\
    \end{tabular}
    \subcaption{\label{table:scaled_pd_c}With $\mathbf{T} = \begin{vmatrix}\nicefrac{2}{3} & \nicefrac{1}{3}\\0 & 1
\end{vmatrix}$}
  \end{subtable}
\end{table}

In order to resolve Scaled-PD, a reward transfer matrix $\mathbf{T} = \begin{vmatrix}\nicefrac{1}{2} & \nicefrac{1}{18}\\\nicefrac{1}{2} & \nicefrac{1}{2}
\end{vmatrix}$ is sufficient, which leaves $\nicefrac{8}{18}$ reward from the row player unaccounted for.
The row player cannot allocate this to themselves, as it would incentivise them to defect, so they must grant this \emph{excess reward} to the column player.
Consequently, both players receive equal reward after transfers (\cref{table:scaled_pd_b}).
However, it can be informative to understand when excess reward exists.
In some sense, this is a fair result, as both players have equal outcomes, but in another sense, it is not fair to the row player, whose rewards are a factor of three greater than the column player's in the original game, and who could therefore argue that they should receive a larger share of the rewards in the transformed game.
The row player might reasonably propose the reward transfer matrix in \cref{table:scaled_pd_c}, where they maintain a three to one ratio of rewards to the column player in the resulting dominant action profile $(D,C)$, yet still greatly benefit the column player, who will receive a payoff close to their best action profile in the unmodified game, at the cost of only an infinitesimal loss of social welfare.

We leave the fairness of the resulting reward distribution as a topic for further work.
However, we have a couple of suggestions to tackle this issue.
First, users can use the general self-interest level to measure the ease of achieving cooperation in a game and utilise additional metrics to quantify the fairness of the post-reward transfer distribution.
A natural choice for a fairness metric may be the Shapley value~\cite{shapley88__a_value_for_n_person_games}, which is intuitively the average marginal contribution an individual makes to their coalition.
This concept has previously been applied to cooperative games~\cite{bachrach10__the_good_the_bad_and_the_cautious}.
Second, to alleviate this issue, the players could create a pseudo-player, called \emph{the lawyer}, who receives any excess rewards and redistributes them in a pre-arranged manner among the players who cooperated.
This enables a player to receive some of their own excess reward without creating an incentive to defect, as doing so makes them ineligible for transfers from the lawyer.
For example, the players could agree on a fairness metric and task the lawyer with distributing the excess rewards to improve this metric.
In the case of Scaled-PD, if $\mathbf{T} = \begin{vmatrix}\nicefrac{1}{2} & \nicefrac{1}{18}\\\nicefrac{1}{2} & \nicefrac{1}{2}
\end{vmatrix}$ the post-transfer reward for $(C,C)$ is $(6,2)$, with 4 excess reward.
Suppose the players task the lawyer with minimising the function $\lvert r_{row} - 2r_{column} \rvert$, then both players would be transferred 2 reward, leaving them with $(8, 4)$.

It is not possible for reward transfer agreements to be rational for all possible beliefs in partial social dilemmas\footnote{We are talking about the subjective probability a player may have over the strategy that their opponent(s) may play, not Bayesian games~\cite{harsanyi04__games_with_incomplete_information_played_by_bayesian_players_iiii} which feature incomplete information.}.
As an example, consider the game of Chicken in \cref{table:sd_b}.
If a player believes that their opponent will cooperate, they anticipate that they can defect and receive a reward of 4.
In this case, they will not be inclined to offer any reward to their opponent, as they do not need to incentivise cooperation.
Therefore, it would not be rational for them to commit to the minimal reward transfer matrix to resolve Chicken, as they would then only receive a reward of 3.

We conclude by contrasting our work against what we believe are the closest alternatives in literature.
Our reward transfer mechanism is the same as that of Gemp et al.~\cite{gemp22__d3c}, where each player receives a convex combination of the rewards of their co-players.
The algorithm they propose, \emph{D3C}, aims to minimise the Price of Anarchy using a novel method for approximating local Nash equilibria, and uses gradient descent to find both the strategies and loss-sharing matrix.
However, by trading off the group loss against minimising reward transfers, D3C may not achieve a social welfare optimum, as suggested by their experiments in stochastic games where D3C fails to achieve the same social welfare as optimal cooperative strategies.
Furthermore, minimising the Price of Anarchy may leave the players with an equilibrium selection problem, similar to the discussion above of Too Many Cooks in Prison with reward exchange (\cref{table:cooks_exchange}).
For example, the Price of Anarchy of Pure Coordination (\cref{table:coordination}) with two players is one, yet both defect and cooperate are rational actions for the players.
Even if all Nash equilibria maximise social welfare, unless there is a unique, pure equilibrium, the players may fail to coordinate.

The approach taken by Schmid et al.~\cite{schmid23__learning_to_participate_through_trading_of_reward_shares} could be adapted to achieve the same mixing of rewards as reward transfer.
However, our approach involves players choosing to donate their rewards to their co-players to incentivise cooperation, instead of actively acquiring a stake in the rewards of the other players via trading.
More importantly, compared to these two papers, we focus on determining the greatest amount of self-interest that the players can maintain while resolving the social dilemma.
This provides our work with a descriptive element, and if reward transfer is expensive, it minimises costs by identifying the smallest amount to be transferred.

From a descriptive perspective, the work of Apt and Schafer~\cite{apt14__selfishness_level_of_strategic_games} is closest, as the selfishness level is also a property of a game quantifying the alignment between individual and group rewards.
However, by offering all players the same additional reward in each action profile, comparable to reward exchange, they cannot leverage the specific incentive structures of a game and may consequently miss more efficient solutions, as we demonstrated in \cref{sec:general_transfer} with Cyclical-3PD (\cref{table:c_pd}).
Additionally, as a prescriptive method, it does not explain where the additional reward originates from.

\section{Experiments}
\label{sec:experiments}
In \cref{sec:game_setup} we introduce four ways to generalise Prisoner's Dilemma, Chicken and Stag Hunt (\cref{table:sd}) to \emph{n}-players.
We find their symmetrical and general self-interest levels (\cref{eqn:symmetrical_self_interest_level,eqn:general_self_interest_level}) as a function of the number of players, and interpret our findings in \cref{sec:results}.
The final \cref{sec:algorithm_results} demonstrates our implementation\footnote{\url{https://github.com/willis-richard/reward_transfer_matrix}} of Algorithm~\ref{alg:mgsd} on two additional games, and provides indicative results for how the computational time increases as a function of the number of players.

\subsection{Graphical dilemmas}
\label{sec:game_setup}
In what follows, we refer to Prisoner's Dilemma, Chicken and Stag Hunt as \emph{base games}.
These base games have payoffs parameterised by two variables, $c$ and $d$.
In all three base games, a player receives an amount $c$ when their opponent cooperates.
Additionally, the player may receive a reward $d$ depending on the joint action between them and their opponent, as detailed in \cref{table:base_payoffs}.
For example, in both Prisoner's Dilemma and Chicken, if a player defects while their opponent cooperates, they will earn a payoff of $c + d$.
Using values of $c=3$ and $d=1$, we obtain the base game payoffs as in \cref{table:sd}.
To satisfy the social dilemma inequalities (\cref{sec:social_dilemma}), we require $c > d$ for Prisoner's Dilemma, and $c > 2d$ for the others.

\begin{table}[ht]
  \centering
  \caption{\label{table:base_payoffs}Base game payoff conditions}
  \begin{tabular}{l|l|l}
    Game & Reward & Condition\\
    \hline
    Prisoner's & c & Opponent cooperated\\
    Dilemma & d & Player defected\\
    \hline
    Chicken & c & Opponent cooperated\\
    & d & Both players took different actions\\
    \hline
    Stag Hunt & c & Opponent cooperated\\
    & d & Both players took the same action\\
  \end{tabular}
\end{table}

We now introduce four novel multi-player normal-form social dilemmas (\cref{sec:social_dilemma}), representing different ways to generalise the two-player base games to \emph{n}-players.
The games are created using a weighted, directed graph with $n$ nodes, each representing a player, combined with a choice of base game.
The base game is played wherever two players share an edge.
The players simultaneously select a single action to be played in all their two-player base games, with each player receiving the weighted rewards where they have an inbound edge.
In this way, the graph determines \emph{who plays with whom}, and the base game determines \emph{what} is played.
Given a graph and a base game, we can compute the full \emph{n}-dimensional payoff matrix.
We refer to the multi-player normal-form social dilemmas constructed in this manner as \emph{graphical dilemmas}.

\begin{figure}[t]
  \centering
  \begin{subfigure}[t]{0.49\textwidth}
    \includegraphics[width=\textwidth]{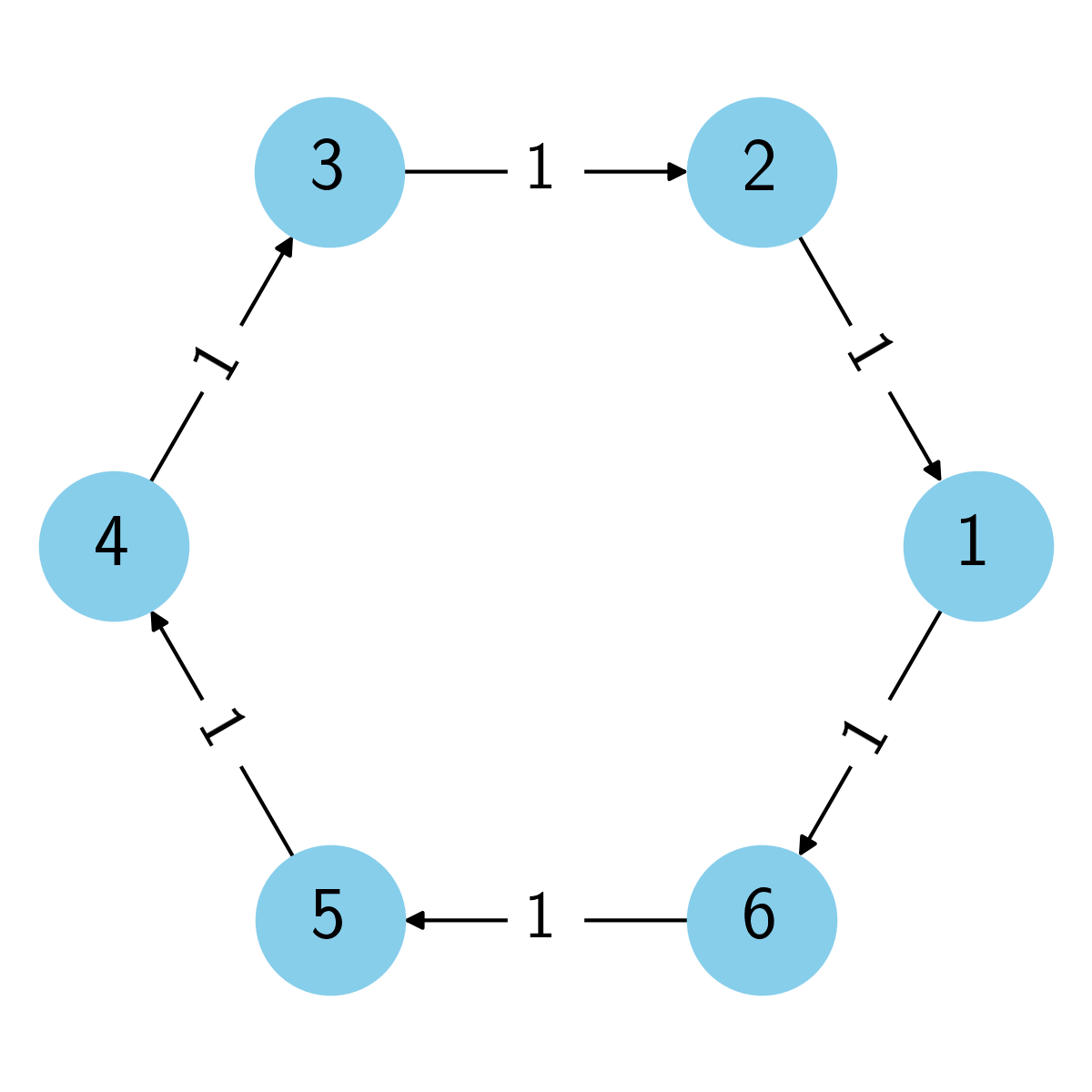}
    \caption{Cyclical: all edges}
    \label{fig:g_b}
  \end{subfigure}
  \hfill
  \begin{subfigure}[t]{0.49\textwidth}
    \includegraphics[width=\textwidth]{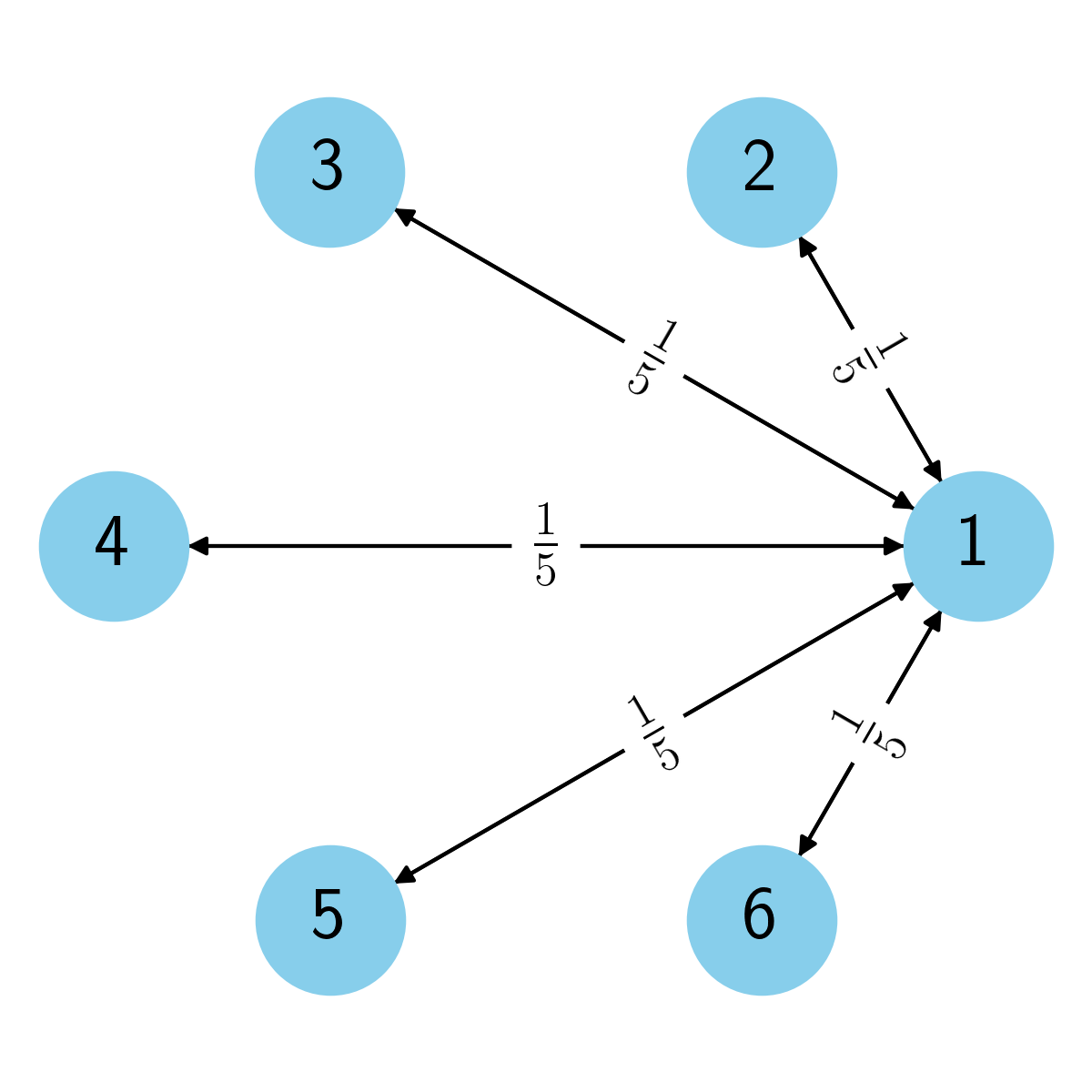}
    \caption{Symmetrical: edges involving player 1}
    \label{fig:g_a}
  \end{subfigure}
  \hfill
  \begin{subfigure}[t]{0.49\textwidth}
    \includegraphics[width=\textwidth]{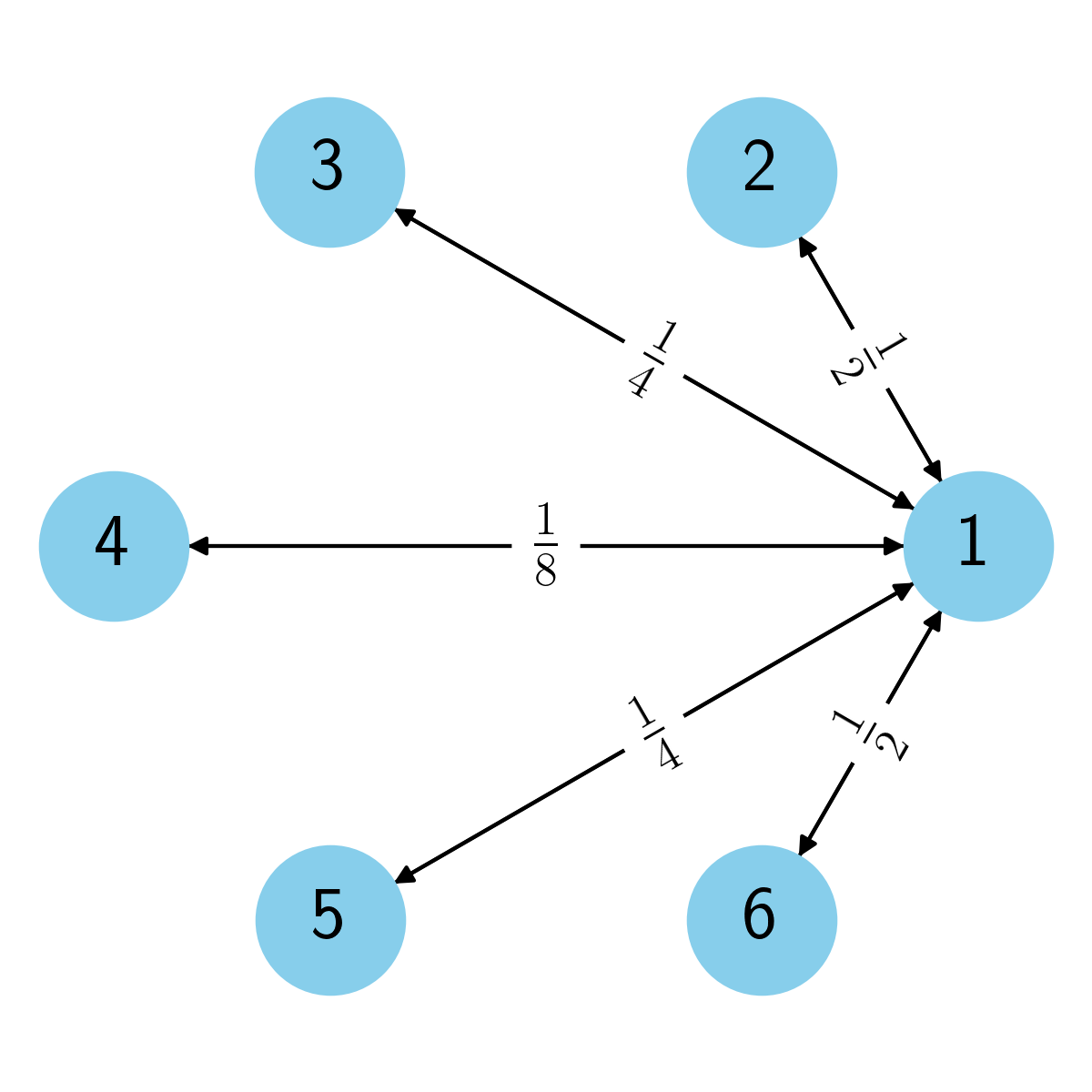}
    \caption{Circular: edges involving player 1}
    \label{fig:g_c}
  \end{subfigure}
  \hfill
  \begin{subfigure}[t]{0.49\textwidth}
    \includegraphics[width=\textwidth]{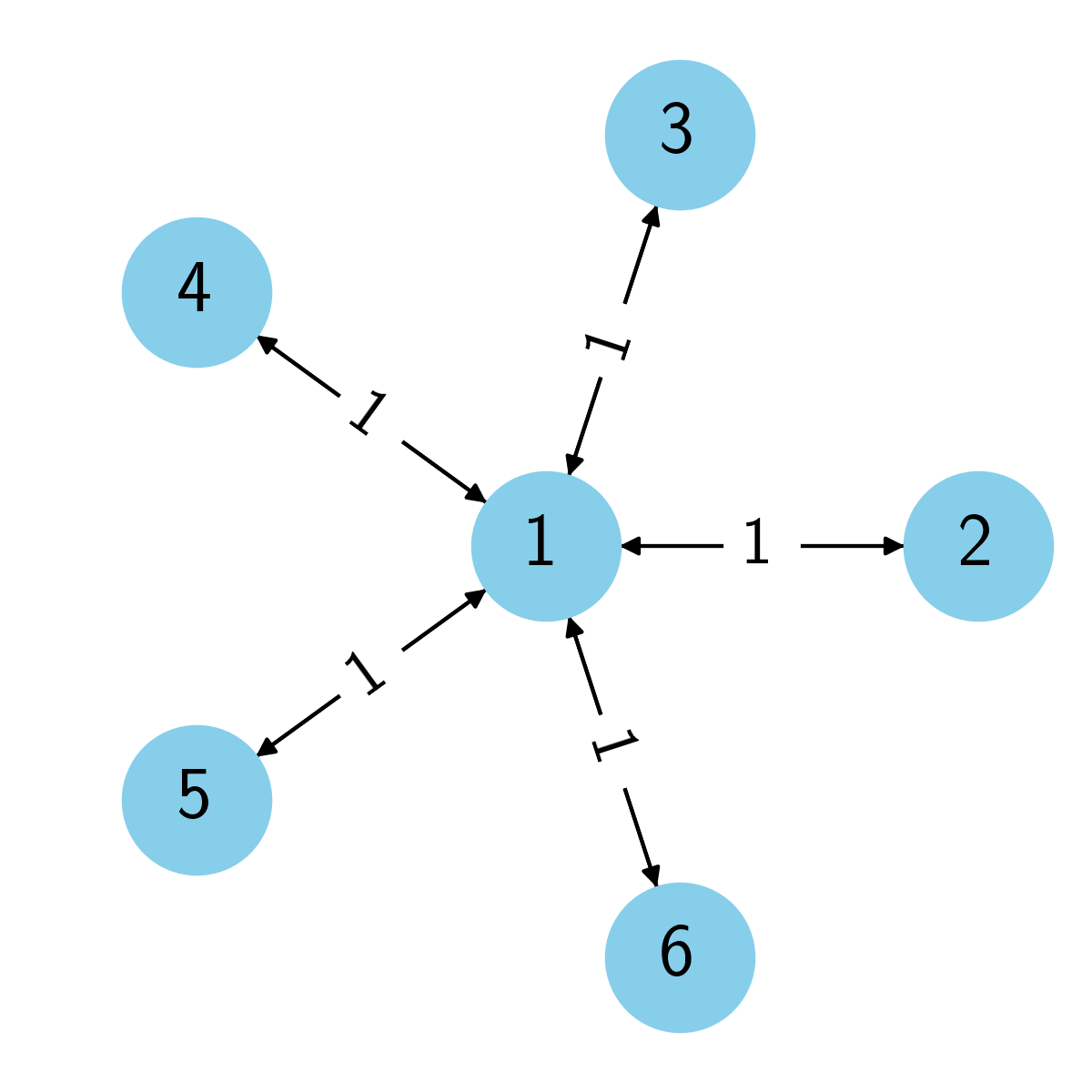}
    \caption{Tycoon: all edges}
    \label{fig:g_d}
  \end{subfigure}
        
  \caption{\label{fig:graphs}Representations of the graphical dilemmas}
\end{figure}

First, we construct Symmetrical and Cyclical graphs, as foreshadowed in \cref{sec:general_transfer}.
We then introduce two additional normal-form social dilemmas, one of which is symmetrical under permutation, while the other has a special player who makes the game asymmetrical.
\cref{fig:graphs} depicts their graphs for $n=6$ players, which are best understood in conjunction with the following descriptions.

\textbf{Cyclical}
Each player $i$ receives a payoff from \cref{table:base_payoffs} with their opponent being the co-player at index $i + 1 \mod n$.
For example, if Chicken is the base game, when player 1 defects and player 2 cooperates, player 1 will receive a reward of $c + d$.
The reward to player 2 depends on the action of player 3.

\textbf{Symmetrical}
Each player plays the base game with every co-player, and their payoffs are weighted by $\frac{1}{n-1}$.
When we aggregate a player's returns, the salient point is the proportion of their opponents who cooperated, which we denote as $n_c$.
\cref{table:symmetrical} summarises their reward given this variable.

\begin{table}[ht]
  \centering
  \caption{\label{table:symmetrical}Rewards using the Symmetrical graph}
  \begin{tabular}{l|l|l}
    Base game & Action & Player reward\\
    \hline
    Prisoner's & C & $c n_c$\\
    Dilemma & D & $d + c n_c$\\
    \hline
    Chicken & C & $c n_c + d (1 - n_c)$\\
    & D & $(c + d)n_c$\\
    \hline
    Stag Hunt & C & $(c + d)n_c$\\
    & D & $c n_c + d (1 - n_c)$
  \end{tabular}
\end{table}

\textbf{Circular}
Similar to the Symmetrical graph, each player interacts with every co-player.
However, this graph additionally models the effect of proximity; players with indices close to each other have a greater impact on each other than those further away.
In the Circular graph, the rewards are scaled by a factor $(\frac{1}{2})^{d(i,j)}$, where $d(i,j) = min(\lvert i - j \rvert, n - \lvert i - j \rvert)$ is the distance between the player at index $i$ and their opponent at index $j$.
Therefore, player 1 has a weighting of $\frac{1}{2}$ when playing with players $n$ and 2, a weighting of $\frac{1}{4}$ when playing with players $n-1$ and 3, and so on.

\textbf{Tycoon}
This graph is asymmetric because it privileges one player over the others.
Player 1 is the tycoon and interacts with each of the other $n-1$ players, who are only connected with the tycoon.
Therefore, the tycoon plays the base game simultaneously with every co-player, using the same action in each game, and receives the sum of their rewards from each interaction.
Players 2 to $n$ receive the base game reward with the tycoon as their opponent.

\subsection{Results of analytical solutions}
\label{sec:results}
The games introduced in \cref{sec:game_setup} are relatively simple to solve because the players are symmetric under permutation.
Consequently, the rows in the minimal reward transfer matrix are permutations of each other.
We were therefore able to determine the inequality constraints necessary to remove any incentives to defect by hand and solve the equations algebraically.
The exception is Tycoon, but here we only have two types of players, so the solution remained tractable.
We now present the results for the games described above, which we have verified using our implementation of Algorithm~\ref{alg:mgsd}.
\cref{table:results} shows the symmetrical and general self-interest levels (\cref{eqn:symmetrical_self_interest_level,eqn:general_self_interest_level}) as functions of the base game parameters (\cref{table:base_payoffs}) and the number of players, $n$, except for Circular where we show the limit that the general self-interest level decreases to as $n$ tends to infinity.

\begin{table}[ht]
  \centering
  \makegapedcells
  \caption{\label{table:results}Self-interest levels of the graphical dilemmas}
  \begin{tabular}{l|l|l|l}
    Graph & Base Game & Symmetrical, $s^*$ & General, $g^*$\\
    \hline
    Cyclical & Prisoner's Dilemma & $\frac{c}{c + d(n-1)}$ & $\frac{c}{c + d}$\\
     & Chicken & $\frac{c-d}{c + d(n-2)}$ & $\frac{c-d}{c}$\\
     & Stag Hunt & $\frac{c-d}{c + d(n-2)}$ & $\frac{c-d}{c}$\\
    \hline
    Symmetrical & Prisoner's Dilemma & $\frac{c}{c + d(n-1)}$ & $\frac{c}{c + d(n-1)}$\\
     & Chicken & $\frac{c-d}{c + d(n-2)}$ & $\frac{c-d}{c + d(n-2)}$\\
     & Stag Hunt & $\frac{c-d}{c + d(n-2)}$ & $\frac{c-d}{c + d(n-2)}$\\
    \hline
    Circular & Prisoner's Dilemma & $\frac{c}{c + d(n-1)}$ & $\lim_{n\to\infty} \frac{c}{c + 4d}$\\
     & Chicken & $\frac{c-d}{c + d(n-2)}$ & $\lim_{n\to\infty} \frac{c-d}{c + 3d}$\\
     & Stag Hunt & $\frac{c-d}{c + d(n-2)}$ & $\lim_{n\to\infty} \frac{c-d}{c + 3d}$\\
    \hline
    Tycoon & Prisoner's Dilemma & $\frac{c}{c + d(n-1)}$ & $\frac{c}{c + d(n-1)}$\\
     & Chicken & $\frac{c-d}{c + d(n-2)}$ & $\frac{c-d)}{c + d(n-2)}$\\
     & Stag Hunt & $\frac{c-d}{c + d(n-2)}$ & $\frac{c-d}{c + d(n-2)}$\\
  \end{tabular}
\end{table}

For all the Graphical dilemmas, the symmetrical self-interest level remains consistent for a given base game.
This uniformity arises because the ratio of any potential gain from defecting compared to the loss of social welfare is constant, since it is determined by the base game rather than the graph.
A symmetrical reward transfer matrix cannot exploit the fact that, for some graphs, a player is impacted to differing degrees by different co-players.
Consequently, the players exchange rewards with players who impact them only minimally, or not at all, which is inefficient.

Additionally, the symmetrical and general self-interest levels are identical with either Chicken or Stag Hunt for all graphs.
In Chicken, we must deter defection against a cooperating opponent, whereas in Stag Hunt, the objective is to remove the incentive for the players to defect against defecting opponents.
Despite this difference, owing to the relative values gained or lost (a personal gain of $d$ at the expense of $c-d$ to the opponent), the transfer structures are the same.

\textbf{Results for Cyclical}
These graphical dilemmas result in sparse minimal reward transfer matrices with only one off-diagonal element being non-zero: that corresponding to the opponent whose cooperative action benefits the row player.
The minimal reward transfer matrix for the Cyclical graph with Prisoner's Dilemma is presented below first, and the second matrix applies to the two other base games.
We use the subscript $P$ to refer to Prisoner's Dilemma and $C,S$ to refer to Chicken and Stag Hunt.

\begin{align*}
  \begin{split}
  \mathbf{T}^*_P = \frac{1}{c+d}
  &\left\lvert
  \begin{array}{C C C C C C C}
    c & d & 0 & \cdots & 0 & 0 & 0\\
    0 & c & d & \cdots & 0 & 0 & 0\\
    0 & 0 & c & \cdots & 0 & 0 & 0\\
    \vdots & \vdots & \vdots & \ddots & \vdots & \vdots & \vdots\\
    0 & 0 & 0 & \cdots & c & d & 0\\
    0 & 0 & 0 & \cdots & 0 & c & d\\
    d & 0 & 0 & \cdots & 0 & 0 & c\\
  \end{array}
  \right\rvert
  \end{split}
  \\[1ex]
  \begin{split}
  \mathbf{T}^*_{C,S} = \frac{1}{c}
  &\left\lvert
  \begin{array}{C C C C C C C}
    c-d & d & 0 & \cdots & 0 & 0 & 0\\
    0 & c-d & d & \cdots & 0 & 0 & 0\\
    0 & 0 & c-d & \cdots & 0 & 0 & 0\\
    \vdots & \vdots & \vdots & \ddots & \vdots & \vdots & \vdots\\
    0 & 0 & 0 & \cdots & c-d & d & 0\\
    0 & 0 & 0 & \cdots & 0 & c-d & d\\
    d & 0 & 0 & \cdots & 0 & 0 & c-d\\
  \end{array}
  \right\rvert
  \end{split}
\end{align*}

Only the first and last three rows and columns are shown to illustrate the scaling pattern.
Thus, in both $\mathbf{T}^*_P$ and $\mathbf{T}^*_{C,S}$, the third row contains a $d$ in the fourth column, and so forth.
The general self-interest level remains constant as the number of players increases because each player is remains impacted by only a single co-player.

\textbf{Results for Symmetrical}
Here, the general self-interest level is equal to the symmetrical self-interest level, and they both tend to zero as $n$ approaches infinity.
Every diagonal element in $\mathbf{T}^*$ is equal to $g^*$ while the off-diagonals can be of any value, within bounds, so long as the columns sum to one.
In this way, each player cares about themselves equal to the general self-interest level, and one minus this value about their co-players in aggregate, but it does not matter which specific combination of co-players.
For illustration, we show two possible minimal reward transfer matrices using Prisoner's Dilemma as the base game.
The first is identical to reward exchange (\cref{sec:exchange}) at the symmetrical self-interest level, the second minimises the total number of transfers by requiring each player to transfer to only one other.

\begin{align*}
  \begin{split}
  \mathbf{T}^*_P = \frac{1}{c+d(n-1)}
  &\left\lvert
  \begin{array}{C C C C C C C}
    c & d & d & \cdots & d & d & d\\
    d & c & d & \cdots & d & d & d\\
    d & d & c & \cdots & d & d & d\\
    \vdots & \vdots & \vdots & \ddots & \vdots & \vdots & \vdots\\
    d & d & d & \cdots & c & d & d\\
    d & d & d & \cdots & d & c & d\\
    d & d & d & \cdots & d & d & c\\
  \end{array}
  \right\rvert
  \end{split}
  \\[1ex]
  \begin{split}
  \mathbf{T}^*_P = \frac{1}{c+d(n-1)}
  &\left\lvert
  \begin{array}{C C C C C C C}
    c & d(n-1) & 0 & \cdots & 0 & 0 & 0\\
    0 & c & d(n-1) & \cdots & 0 & 0 & 0\\
    0 & 0 & c & \cdots & 0 & 0 & 0\\
    \vdots & \vdots & \vdots & \ddots & \vdots & \vdots & \vdots\\
    0 & 0 & 0 & \cdots & c & d(n-1) & 0\\
    0 & 0 & 0 & \cdots & 0 & c & d(n-1)\\
    d(n-1) & 0 & 0 & \cdots & 0 & 0 & c\\
  \end{array}
  \right\rvert
  \end{split}
\end{align*}

Despite our ability to form a sparse transfer matrix, the scenario differs significantly from the Cyclical graph.
Here, the players are fully connected and equally impacted by all co-players, so their influence on any given player diminishes as the number of players increases.
Consequently, while a player sacrifices $d$ by cooperating, the benefit of their cooperation is dispersed among each co-player, who individually receive only $\frac{c}{n-1}$.
Thus, each co-player has less reward to offer in return for the cooperation.
It is therefore necessary that the self-interest of the players also diminishes as $n$ increases, thus reducing the benefit from defecting.
This analysis applies similarly for the Chicken and Stag Hunt base games.

\textbf{Results for Circular}
The Circular graphs have a general self-interest level that decreases to the value given in \cref{table:results}.
At this point, the minimal transfer matrices have the following forms.

\begin{align*}
  \begin{split}
  \mathbf{T}^*_P = \frac{1}{c + 4d}
  &\left\lvert
  \begin{array}{C C C C C C C}
    c & 2d & 0 & \cdots & 0 & 0 & 2d \\
    2d & c & 2d & \cdots & 0 & 0 & 0 \\
    0 & 2d & c & \cdots & 0 & 0 & 0\\
    \vdots & \vdots & \vdots & \ddots & \vdots & \vdots & \vdots\\
    0 & 0 & 0 & \cdots & c & 2d & 0\\
    0 & 0 & 0 & \cdots & 2d & c & 2d\\
    2d & 0 & 0 & \cdots & 0 & 2d & c
  \end{array}
  \right\rvert
  \end{split}
  \\[1ex]
  \begin{split}
  \mathbf{T}^*_{C,S} = \frac{1}{c + 3d}
  &\left\lvert
  \begin{array}{C C C C C C C}
    c-d & 2d & 0 & \cdots & 0 & 0 & 2d \\
    2d & c-d & 2d & \cdots & 0 & 0 & 0 \\
    0 & 2d & c-d & \cdots & 0 & 0 & 0\\
    \vdots & \vdots & \vdots & \ddots & \vdots & \vdots & \vdots\\
    0 & 0 & 0 & \cdots & c-d & 2d & 0\\
    0 & 0 & 0 & \cdots & 2d & c-d & 2d\\
    2d & 0 & 0 & \cdots & 0 & 2d & c-d
  \end{array}
  \right\rvert
  \end{split}
\end{align*}

Despite every player being impacted by the actions of all co-players, they only need to transfer rewards to those who impact them most, because the other players are induced to cooperate by their respective neighbours.
This forms the most efficient transfer structure; players are compensated by those most impacted by them, which minimises the proportion of rewards that they need to transfer, and provides a lower bound on the self-interest of this game.

\textbf{Results for Tycoon}
This graph gives rise to reward transfer matrices with excess rewards (introduced in \cref{sec:method_discussion}).
They specify transfers of the following forms.

\begin{align*}
  \begin{split}
  \mathbf{T}_P = \frac{1}{c + d(n-1)}
  &\left\lvert
  \begin{array}{C C C C C C C}
    c & d & d & \cdots & d & d & d\\
    d & c & 0 & \cdots & 0 & 0 & 0\\
    d & 0 & c & \cdots & 0 & 0 & 0\\
    \vdots & \vdots & \vdots & \ddots & \vdots & \vdots & \vdots\\
    d & 0 & 0 & \cdots & c & 0 & 0\\
    d & 0 & 0 & \cdots & 0 & c & 0\\
    d & 0 & 0 & \cdots & 0 & 0 & c\\
  \end{array}
  \right\rvert
  \end{split}
  \\[1ex]
  \begin{split}
  \mathbf{T}_{C,S} = \frac{1}{c + d(n-2)}
  &\left\lvert
  \begin{array}{C C C C C C C}
    c-d & d & d & \cdots & d & d & d\\
    d & c-d & 0 & \cdots & 0 & 0 & 0\\
    d & 0 & c-d & \cdots & 0 & 0 & 0\\
    \vdots & \vdots & \vdots & \ddots & \vdots & \vdots\\
    d & 0 & 0 & \cdots & c-d & 0 & 0\\
    d & 0 & 0 & \cdots & 0 & c-d & 0\\
    d & 0 & 0 & \cdots & 0 & 0 & c-d\\
  \end{array}
  \right\rvert
  \end{split}
\end{align*}

Note that only the first row, representing the tycoon, sums to 1.
This is because the tycoon is the limiting factor, who must incentivise all $n-1$ co-players to cooperate.
When Tycoon is paired with Prisoner's Dilemma, in order to ensure that cooperation is dominant for players $i = 2, ..., n$, their self-interest cannot exceed $\frac{c}{d}$ times the proportion of reward offered by the tycoon.
Collectively, the other players incentivise the tycoon to cooperate, which they achieve by each offering the tycoon $\frac{d}{c + d(n-1)}$.
This leaves the non-tycoon players $\frac{d(n-2)}{c + d(n-1)}$ excess rewards, which they can freely distribute to any player other than themselves.
In fact, there is no requirement for each player to transfer exactly $\frac{d}{c + d(n-1)}$ to the tycoon, the salient point is that the tycoon receives at least $\frac{d(n-1)}{c + d(n-1)}$ in total.
We provide two example minimal reward transfer matrices, the first maximising the reward to the tycoon, the second distributing all the surplus reward equally among the other players.

\begin{align*}
  \begin{split}
  \mathbf{T}^*_P = \frac{1}{c+d(n-1)}
  &\left\lvert
  \begin{array}{C C C C C C C}
    c & d & d & \cdots & d & d & d\\
    d(n-1) & c & 0 & \cdots & 0 & 0 & 0\\
    d(n-1) & 0 & c & \cdots & 0 & 0 & 0\\
    \vdots & \vdots & \vdots & \ddots & \vdots & \vdots & \vdots\\
    d(n-1) & 0 & 0 & \cdots & c & 0 & 0\\
    d(n-1) & 0 & 0 & \cdots & 0 & c & 0\\
    d(n-1) & 0 & 0 & \cdots & 0 & 0 & c\\
  \end{array}
  \right\rvert
  \end{split}
  \\[1ex]
  \begin{split}
  \mathbf{T}^*_P = \frac{1}{c+d(n-1)}
  &\left\lvert
  \begin{array}{C C C C C C C}
    c & d & d & \cdots & d & d & d\\
    d & c & d & \cdots & d & d & d\\
    d & c & d & \cdots & d & d & d\\
    \vdots & \vdots & \vdots & \ddots & \vdots & \vdots\\
    d & d & d & \cdots & c & d & d\\
    d & d & d & \cdots & d & c & d\\
    d & d & d & \cdots & d & d & c\\
  \end{array}
  \right\rvert
  \end{split}
\end{align*}

For this graph, the general self-interest level is equivalent to the symmetrical self-interest level, as is evident from the second matrix above being identical to reward exchange.
However, as discussed, there is a range of possible solutions, leaving different players better or worse off after reward transfers.

\textbf{Discussion}
By analysing the minimal transfer matrix, we gain insights into the game structure.
As observed in the Cyclical and Circular graphs, players need only provide incentives to those who impact them most strongly.
This may be indicative of a broader trend across games, where minimal reward transfer matrices tend to be sparse.
Such sparsity would imply that players typically need to transfer rewards to only a few others, minimising the total number of transfers. However, we leave this observation for further research.

\subsection{Results of approximation solutions}
\label{sec:algorithm_results}
\textbf{Arbitrary social dilemma}
To demonstrate \cref{alg:mgsd}, we construct a three-player normal-form social dilemma as depicted in \cref{table:arbitrary}.
Here, there is no pattern to the change in payoffs, so it is tedious to solve by hand.
The minimal reward transfer matrix found by our implementation, rounded to three decimal places, is as follows.

\begin{table}[ht]
  \centering
  \caption{\label{table:arbitrary}Arbitrary social dilemma}
  \begin{subtable}{0.49\linewidth}
    \centering
    \makegapedcells
    \begin{tabular}{c|cc}
      & $C$ & $D$\\
      \hline
      $C$ & $9,6,7$ & $2,9,7$\\
      $D$ & $8,4,8$ & $3,2,1$\\
    \end{tabular}
    \subcaption{Player 3 cooperates}
  \end{subtable}
  \begin{subtable}{0.49\linewidth}
    \centering
    \makegapedcells
    \begin{tabular}{c|cc}
      & $C$ & $D$\\
      \hline
      $C$ & $1,6,12$ & $0,5,2$\\
      $D$ & $8,2,8$ & $1,2,0$\\
    \end{tabular}
    \subcaption{Player 3 defects}
  \end{subtable}
\end{table}

\begin{equation}
  \nonumber
  \mathbf{T}^* =
  \begin{vmatrix}
    0.487 & 0.209 & 0.304\\
    0.426 & 0.487 & 0.087\\
    0.426 & 0.087 & 0.487
  \end{vmatrix}
\end{equation}

The symmetrical self-interest level for this game is $s^* = 0.364$, while $g^* = 0.487$.
We verify the transfer matrix is optimal by inspecting the transformed game, displayed in \cref{table:arbitrary_ptr} with values rounded to 2 decimal places.
The action profile (C,C,C) is dominant, with each player having at least one action profile where they are indifferent between D or C, indicating that the coefficients of $\mathbf{T}^*$ are at their limits.
As the minimal reward transfer matrix is not unique for this game, we have opted to present the one with the highest entropy.
In fact, players 2 and 3 have excess rewards, which can be assessed by either systematically reducing their off-diagonal coefficients and checking that cooperation remains dominant, or relaxing the constraint that the rows sum to one in Algorithm~\ref{alg:mgsd} so that they sum to \emph{at most} one.

\begin{table}[ht]
  \centering
  \caption{\label{table:arbitrary_ptr}Arbitrary social dilemma post transfers}
  \begin{subtable}{0.49\linewidth}
    \centering
    \makegapedcells
    \begin{tabular}{c|cc}
      & $C$ & $D$\\
      \hline
      $C$ & $9.92, 5.41, 6.67$ & $7.79, 5.41, 4.80$\\
      $D$ & $9.01, 4.31, 6.68$ & $2.74, 1.69, 1.57$\\
    \end{tabular}
    \subcaption{Player 3 cooperates}
  \end{subtable}
  \begin{subtable}{0.49\linewidth}
    \centering
    \makegapedcells
    \begin{tabular}{c|cc}
      & $C$ & $D$\\
      \hline
      $C$ & $8.16, 4.17, 6.67$ & $2.98, 2.61, 1.41$\\
      $D$ & $8.16, 3.34, 6.50$ & $1.34, 1.18, 0.48$\\
    \end{tabular}
    \subcaption{Player 3 defects}
  \end{subtable}
\end{table}

\textbf{Functional social dilemma}
In order to assess the practical runtime of computing a minimal reward transfer matrix, we designed a functional game and measured the time our solver required to find a solution as the number of players increased.
The game operates as follows:
\begin{itemize}
    \item The social welfare is defined as $SW(\vv{a}) = -\frac{c}{n}n_c^2 + 2cn_c$, where $n$ is the number of players, $n_c$ represents the number of players cooperating, and $c$ is a constant that scales the reward.
    \item Each player has a weight equal to their player index, one for player 1, two for player 2, and so on. Choosing defect doubles their weight.
    \item A player receives a proportion of the social welfare equal to their effective weight divided by the total player weight.
\end{itemize}

\begin{figure}[htbp]
\centering
\includegraphics[width=200px]{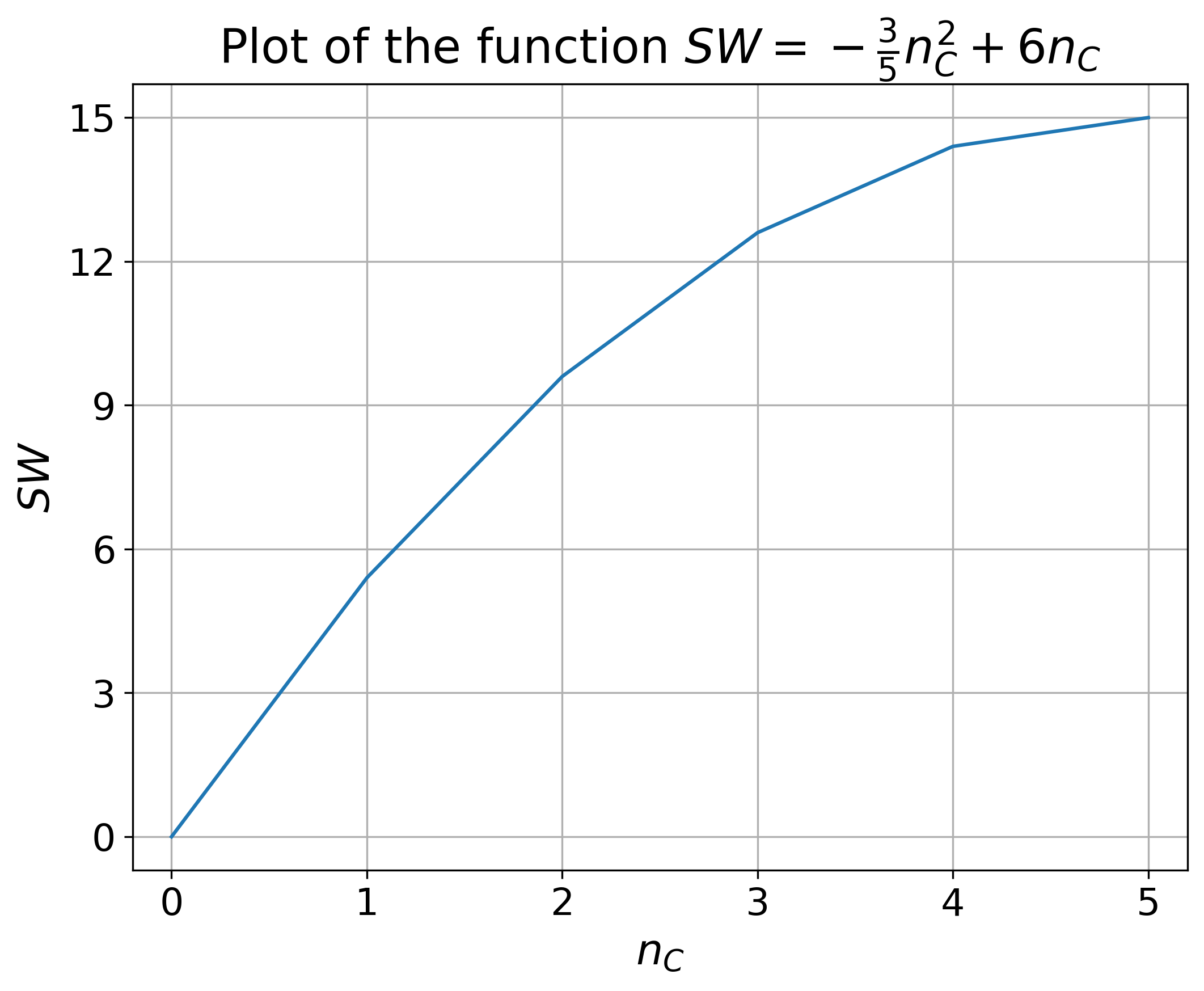}
\caption{\label{fig:fsd}Social welfare of the Functional social dilemma}
\end{figure}

Figure~\ref{fig:fsd} shows the social welfare for the Functional social dilemma with $c=3$ and $n=5$.
This game resembles a multi-player version of Chicken, in that players can enhance their rewards by defecting, as long as the majority of their co-players are cooperating.
Due to the asymmetry of the players, they have different incentives to defect, in contrast to the graphical dilemmas.
Our implementation\footnote{\url{https://github.com/willis-richard/reward_transfer_matrix/}} of \cref{alg:mgsd} utilises a linear program solver from the Python library SciPy\footnote{\url{https://docs.scipy.org/doc/scipy/reference/generated/scipy.optimize.linprog.html}}, which employs a simplex method~\cite{dantzig63__linear_programming_and_extensions} from the HiGHS library\footnote{\url{https://highs.dev/}}.
We find the following unique minimal reward transfer matrix for this configuration of Functional social dilemma, presented to 3 decimal places.

\begin{equation}
  \nonumber
  \mathbf{T}^* =
  \begin{vmatrix}
    0.471 & 0.395 & 0.000 & 0.000 & 0.134\\
    0.220 & 0.471 & 0.000 & 0.309 & 0.000\\
    0.418 & 0.000 & 0.471 & 0.035 & 0.076\\
    0.000 & 0.067 & 0.342 & 0.471 & 0.121\\
    0.089 & 0.269 & 0.061 & 0.110 & 0.471\\
  \end{vmatrix}
\end{equation}

We provide an indication of how computational time might scale in practice by computing a minimal reward transfer matrix across a range of values of $n \in [8,17]$.
Our implementation achieves the times shown in \cref{table:solver}  when run on a commercial laptop.
The solving time increases approximately threefold for each additional player, suggesting that the complexity is $\approx O(3^n)$, in line with our expectations of exponential complexity from the analysis in \cref{sec:method_algorithm}.
Consequently, the solving time will be prohibitively long for large numbers of players.
However, we note that the size of representing the game in matrix form is of order $O(2^nn)$, which also becomes prohibitively large.
For instance, a normal-form social dilemma with $n=30$ players requires around 30 billion floating point numbers to represent.

\begin{table}[ht]
  \centering
  \caption{\label{table:solver}Solving time for the Functional social dilemma}
    \begin{tabular}{cc|cc}
      $n$ & time (s) & $n$ & time(s)\\
      \hline
      8 & 0.02 & 13 & 1.68\\
      9 & 0.04 & 14 & 4.06\\
      10 & 0.10 & 15 & 11.04\\
      11 & 0.24 & 16 & 33.68\\
      12 & 0.71 & 17 & 85.77\\
    \end{tabular}
\end{table}

\section{Conclusion}
\label{sec:conclusion}
In this paper, we considered the problem of designing incentives for collective action in social dilemmas.
We introduced a mechanism whereby agents enter into a binding agreement to transfer portions of their future rewards to others.
In doing so, agents who were previously self-interested now have an incentive to prefer that their partners achieve high rewards, thereby making prosocial actions more appealing.
This mechanism serves two purposes: first, to establish two metrics that describe the alignment between what is individually rational, and what is collectively rational; and second, to propose a solution to social dilemmas by making cooperation the rational choice for all agents.
Beyond a solution to resolve social dilemmas, potential applications of these metrics are to be found in the field of mechanism design, where we can assess the impact on collaborative behaviour of modifications to models of environments.

The first metric, termed the symmetrical self-interest level (\cref{sec:exchange}), posits that all players exchange the same proportion of their rewards with each other and is comparable to other metrics in the literature~\cite{apt14__selfishness_level_of_strategic_games}.
It denotes the maximum degree of self-interest that players can retain while resolving a social dilemma under such an exchange scheme.
The advantage of this metric is its simplicity of computation, as it involves only one parameter.
The second metric, named the general self-interest level (\cref{sec:general_transfer}), relaxes the requirement that all players exchange the same proportion of rewards and permits players to transfer rewards in an unrestricted manner.
This flexibility can lead to more efficient solutions, where players transfer less reward by strategically considering which specific co-players they would benefit most from influencing.
Consequently, the optimal transfer arrangement supporting the general self-interest level of the game can provide insights into the underlying game structure.

We subsequently introduced novel multi-player normal-form social dilemmas and presented results for them (\cref{sec:results}).
Our results (\cref{sec:results}) suggest that the minimal transfer matrix may have a tendency to be sparse, which limits the number of transfers required as the number of players increases.
Lastly, we developed an algorithm (\cref{sec:method_algorithm}) to determine the general self-interest level and identify an optimal scheme of transfers.
This algorithm employs a linear program to compute the minimal reward transfer matrix.
To validate our implementation and assess the practical runtime relative to the number of players, we demonstrated its application on two additional games (\cref{sec:algorithm_results}).

The limiting values quantified by the metrics are sufficient to ensure that cooperation is the dominant strategy.
We have presented our method for social dilemma normal-form games with a single social optima under mutual cooperation.
Nevertheless, we contend that our approach can be generalised to encompass various definitions of social dilemmas, as illustrated in \cref{sec:method_discussion}.
In this case, if there are multiple action profiles that maximise social welfare, the players will need to agree in advance which social optimum to make dominant.
Once selected, the process to render that action profile dominant will mirror the procedure in  \cref{alg:mgsd}, albeit targeted at a different action profile.

Directions for future work include developing methods to compute the metrics for other classes of games, such as Markov games, Bayesian games and robust games.
Recent work~\cite{hughes18__inequity_aversion_improves_cooperation_in_intertemporal_social_dilemmas} has formalised social dilemmas in terms of partially observable Markov games.
Such \emph{sequential social dilemmas} model more aspects of real world social dilemmas~\cite{leibo17__multiagent_reinforcement_learning_in_sequential_social_dilemmas}, including cooperativeness as a graded quantity, agents having only partial information about the state of the world, and decisions being temporally extended from their consequences.
Due to their extensive state space, these games are computationally intractable; they require optimisation algorithms to determine effective strategies.
See~\cite{koster20__modelfree_conventions_in_multi-agent_reinforcement_learning_with_heterogeneous_preferences} for a detailed review of the challenges facing learning agents in a sequential social dilemma.
A method to determine the optimal symmetrical reward transfer matrix (reward exchange) has been proposed~\cite{willis23__resolving_social_dilemmas_through_reward_transfer_commitments}, but it requires adaptation to find the optimal general reward transfer matrix.
We also discussed the need for metrics that assess the fairness of post-transfer game rewards compared to the unmodified game rewards in \cref{sec:method_discussion}.

Limitations of our approach include a requirement that the players are instrumentally rational, have access to a common currency that is equally valued by all players, and the capability to make commitments or enter binding agreements.
If some or all agents are irrational, then the predictions from game theory may not apply.
Though it may be possible to model the irrationality of the players and adjust the game rewards accordingly, given sufficient information, in the general case non-rationality presents significant challenges for any method.
Furthermore, the terms of the optimal reward transfers are not guaranteed to be rational for all possible player beliefs, which could limit the applicability of our solutions.
Finally, if a general normal-form social dilemma contains no exploitable structure enabling analytical results, we must use an optimisation method to find the minimal reward transfer matrix.
Our implementation traverses all action profiles, which grow exponentially with the number of players in a game, becoming infeasible for large numbers of players.

It is worth noting that while those engaging in mutual cooperation benefit from doing so, it is not always the case that this is better for society if there are negative impacts on those excluded, such as in cases of collusion between firms.
See~\cite{dafoe20__open_problems_in_cooperative_ai} for a more detailed overview of the potential harms of cooperation.

\section*{Competing interests}
The authors have no relevant financial or non-financial interests to disclose beyond the grant information detailed on the title page.

\bibliography{library}

\end{document}